\documentclass[fleqn,usenatbib]{mnras}
\hypersetup{linkcolor=red}          
\usepackage{multirow}
\usepackage[dvipsnames]{xcolor}
\usepackage{newtxtext,newtxmath}
\usepackage{orcidlink}
\usepackage{tabularx}
\usepackage{array}
\usepackage{comment}
\newcolumntype{C}{>{\centering\arraybackslash}p{0.085\textwidth}}
\newcolumntype{Y}{>{\centering\arraybackslash}X}

\def\skirt{{\sc skirt} }
\def\colibre{{\sc colibre} }
\def\cosk{{\sc colibre-skirt} }

\def\equationautorefname~#1\null{Eq.~#1\null}

\title[Local COLIBRE LFs]{Galaxy luminosity functions from far-UV to submillimetre at $z=0$ in the COLIBRE simulations}

\author[S. Lu et al.]
{Shengdong Lu\orcidlink{0000-0002-6726-9499}$^{1}$\thanks{E-mail: \url{shengdong.lu@durham.ac.uk}}, 
Carlos S. Frenk\orcidlink{0000-0002-2338-716X}$^{1}$, 
Cedric G. Lacey\orcidlink{0000-0001-9016-5332}$^{1}$, 
Andrea Gebek\orcidlink{0000-0002-0206-8231}$^{2}$,
Joop Schaye\orcidlink{0000-0002-0668-5560}$^{3}$, 
\and
Shaun Cole\orcidlink{0000-0002-5954-7903}$^{1}$, 
Sownak Bose\orcidlink{0000-0002-0974-5266}$^{1}$, 
Nick Andreadis\orcidlink{0009-0001-9915-6325}$^{2}$,
Maarten Baes\orcidlink{0000-0002-3930-2757}$^{2}$,
Alejandro Benítez-Llambay$^{4}$,
\and
Evgenii Chaikin$^{1,3}$,
Robert A. Crain$^{5}$,
Anna Durrant$^{5}\orcidlink{0009-0003-9672-2170}$,
Filip Huško$^{3}$,
Sylvia Ploeckinger$^{6}$,
\and
Alexander J. Richings$^{7,8}$,
Matthieu Schaller\orcidlink{0000-0002-2395-4902}$^{3,9}$, 
James W. Trayford$^{10}$
\\
\\
$^{1}$Institute for Computational Cosmology, Department of Physics, University of Durham, South Road, Durham, DH1 3LE, UK\\
$^{2}$Department of Physics and Astronomy, Universiteit Gent, Proeftuinstraat 86 N3, B-9000 Ghent, Belgium\\
$^{3}$Leiden Observatory, Leiden University, PO Box 9513, 2300 RA Leiden, the Netherlands\\
$^{4}$Dipartimento di Fisica G. Occhialini, Universit\`a degli Studi di Milano Bicocca, Piazza della Scienza, 3 I-20126 Milano MI, Italy\\
$^{5}$Astrophysics Research Institute, Liverpool John Moores University, 146 Brownlow Hill, Liverpool L3 5RF, UK\\
$^{6}$Department of Astrophysics, University of Vienna, Türkenschanzstrasse 17, A-1180 Vienna, Austria\\
$^{7}$Centre for Data Science, Artificial Intelligence and Modelling, University of Hull, Cottingham Road, Hull, HU6 7RX, UK\\
$^{8}$E. A. Milne Centre for Astrophysics, University of Hull, Cottingham Road, Hull, HU6 7RX, UK\\
$^{9}$Lorentz Institute for Theoretical Physics, Leiden University, PO Box 9506, 2300 RA Leiden, the Netherlands\\
$^{10}$Centre for Astrophysics Research, Department of Physics, Astronomy and Mathematics, University of Hertfordshire, College Lane, Hatfield, AL10 9AB, UK
}
\date{}
\pubyear{2026}

\begin{document}
\label{firstpage}
\pagerange{\pageref{firstpage}--\pageref{lastpage}}
\maketitle
\begin{abstract}
We present predictions from the {\sc colibre} cosmological hydrodynamical simulations of galaxy formation for the present-day galaxy luminosity functions (LFs) at wavelengths ranging from the far-ultraviolet (FUV) to the submillimetre. The simulations are post-processed with the radiative transfer code {\sc skirt}, accounting for dust attenuation and emission using the distribution and properties of dust grains predicted directly by {\sc colibre}. Results from simulations varying in mass resolution by a factor of $\sim 10^2$ ($\sim 10^5 - 10^7\,\mathrm{M_{\odot}}$) show very good convergence over most luminosity ranges. The {\sc colibre-skirt} LFs match the data remarkably well from the FUV to $3.4\,\micron$ and also in the far-infrared and submillimetre wavelength range ($70-850\,\micron$). In the mid-infrared (MIR; $8-24\,\micron$), {\sc colibre-skirt} matches the data well at low luminosities but significantly underpredicts the luminosities of MIR-bright galaxies, with the discrepancy increasing towards longer wavelengths. The total infrared LF, obtained by integrating the spectral energy distributions over $8-1000\,\micron$, also matches observations well at the faint end but underpredicts the number of very bright galaxies. The unprecedented agreement at all other wavelengths indicates that {\sc colibre}, coupled with this calibration-free {\sc skirt} post-processing framework, successfully predicts the properties of stellar populations at the present day and the amount and distribution of interstellar dust.
\end{abstract}


\section{Introduction}
\label{sec:introduction}
Cosmological hydrodynamical simulations are a powerful tool for studying galaxy formation and evolution. The growth of dark matter structures is modelled in a cosmological framework, together with more complex baryonic processes, such as gas cooling, star formation and stellar and active galactic nuclei (AGN) feedback, in a self-consistent manner. Such simulations provide a physically motivated link between the initial conditions in the early Universe and the properties of galaxies observed today. Over the past decade, advances in numerical methods, subgrid physics models and computational power have enabled state-of-the-art cosmological hydrodynamical simulations to reproduce many key observed galaxy statistics across cosmic time, including, for example, stellar mass functions (SMF), star formation rates (SFR), galaxy morphologies and chemical enrichment patterns. These and other properties are reproduced to varying degrees in simulations such as EAGLE \citep{Schaye2015,Crain2015}, Horizon-AGN \citep{Dubois2016,Kaviraj2017}, FABLE \citep{Henden2018}, IllustrisTNG \citep{Nelson2019}, SIMBA \citep{Dave2019}, FLAMINGO \citep{Schaye2023}, MillenniumTNG \citep{Pakmor2023} and in {\sc colibre}, the most recent simulations \citep{Schaye2026,Chaikin2026a}. As a result, cosmological simulations have become an essential tool for interpreting multi-wavelength observations and for testing theoretical models of galaxy formation within a cosmological context.

Despite their significant successes, cosmological simulations remain limited by finite resolution and the need for subgrid models, and therefore require calibration of some of the subgrid model parameters by reference to observations. The stellar mass function (SMF) is a natural choice for a calibration statistic because the stellar mass of galaxies is a direct output of most simulations and can be inferred observationally from fitting stellar population models to broadband spectral energy distributions (SEDs) of galaxies, albeit with several assumptions. Reproducing the observed SMF at $z \approx 0$ has become a standard benchmark for tuning key physical prescriptions in simulations, such as the star formation efficiency and stellar and AGN feedback (e.g. \citealt{Crain2015,Nelson2019,Kugel2023}). This commonly adopted approach is effectively a form of \emph{inverse modelling}, in which observed stellar masses derived from SED fitting are compared directly to simulated ones. However, observationally inferred stellar masses rely on uncertain assumptions about star formation histories, metallicity, dust attenuation and the initial mass function of stars (IMF; e.g. \citealt{Mitchell2013,Lo_Faro2017}). 

Once the model parameters have been calibrated, the model can be evaluated and tested against observations through \emph{forward modelling} whereby simulated galaxy properties are used to predict directly observable quantities, such as luminosities, enabling a direct comparison with observational data. While forward modelling is also subject to uncertainties through the choice of stellar population synthesis models (with fixed IMF choices and binary fractions) and dust modelling, it uses more information from the simulations in a self-consistent way (e.g. the star formation history, metallicity and dust content of each individual galaxy). Furthermore, comparisons based on luminosity-related properties can naturally incorporate observational selection effects and survey bandpasses, thereby providing a powerful test of the modelling behind the simulated galaxy populations.

Luminosity functions (LFs) therefore provide an important diagnostic of the realism with which physical processes are modelled in cosmological simulations. Luminosities measured at different wavelengths trace distinct physical components and reflect distinct processes within galaxies, therefore offering multi-dimensional constraints on the models. For example, ultraviolet (UV) emission primarily traces recent star formation and the presence of young, massive stars (e.g. \citealt{Kennicutt2012,Calzetti2013}); optical and near-infrared (NIR) light is more closely related to the accumulated stellar mass (e.g. \citealt{Bell2001,Zibetti2009}); mid- and far-infrared (MIR and FIR) emission mainly originates from dust heated by starlight, providing insight into dust content and obscured star formation (e.g. \citealt{Draine2007a,Draine2007b,Calzetti2007}); and submillimetre emission traces the properties of the cold dust reservoir (e.g. \citealt{Hildebrand1983,Casey2014}). Joint modelling of these LFs therefore allows robust and mutually consistent constraints on star formation, dust attenuation and dust emission.

To compute luminosities and, further, LFs from cosmological simulations, one must translate the simulated physical properties of galaxies (e.g. stellar mass and SFR) into observable emission using appropriate forward modelling techniques. These include stellar population synthesis (SPS) methods, such as BC03 \citep{Bruzual2003} and {\sc fsps} \citep{Conroy2009,Conroy2010}, as well as the more sophisticated radiative transfer methods that explicitly model the interaction between starlight and dust, such as {\sc grasil} \citep{Silva1998}, {\sc sunrise} \citep{Jonsson2006}, {\sc skirt} \citep{Baes2011,Camps2015,Camps2020} and {\sc powderday} \citep{Narayanan2021}. For example, \citet{Trayford2015} employed stellar population synthesis to study the colours and broadband galaxy LFs in the EAGLE simulations. \citet{Vogelsberger2020b} used {\sc fsps} to model the intrinsic stellar emission and, for their most sophisticated dust treatment, performed dust radiative transfer calculations with a modified version of {\sc skirt} to predict high-redshift UV LFs for galaxies in the IllustrisTNG simulations. \citet{Baes2020} used infrared fluxes computed with {\sc skirt} post-processing by \citet{Camps2018} to study the infrared (IR) galaxy LFs of EAGLE. Also using {\sc skirt}, \citet{Trcka2022} and, more recently, \citet{Gebek2024} investigated the colours and broadband LFs of local galaxies in the IllustrisTNG simulation, focusing on TNG50 and TNG100, respectively. Similar studies also include, for example, those of \citet{Camps2016,Trayford2017,Shen2020,Trcka2020}.

In the earlier studies mentioned above, the spatial distribution of dust, which is crucial for attenuation in the optical and UV bands and for emission in the IR, is not predicted by the simulation itself, but treated in an approximate manner. For example, \citet{Camps2016} assumed a constant dust-to-metal ratio for gas particles meeting certain temperature criteria or associated with ongoing star formation in the EAGLE simulations (see also \citealt{Gebek2024} for IllustrisTNG). This simplifying assumption is mainly adopted because large-volume cosmological simulations such as EAGLE and IllustrisTNG do not explicitly model the cold phase of the interstellar medium (ISM), and therefore do not self-consistently trace the formation, growth and destruction of dust within galaxies.

{\sc colibre} \citep{Schaye2026,Chaikin2026a} is a new state-of-the-art cosmological simulation suite, the first generation of large-volume simulations that explicitly models the multiphase ISM, including a cold ISM component, and simultaneously follows the formation, growth, and destruction of dust in a self-consistent manner. In this series of three papers, we aim comprehensively to study the LFs of the new \colibre simulations across wavelengths from the far-ultraviolet (FUV) to the submillimetre, and over redshifts from $z = 0$ to the early Universe (up to $z = 20$). In this paper (Paper~I), we focus on the local ($z = 0$) LFs from the FUV to the submillimetre and compare them with local observational data. In Paper~II (Lu et al. in prep.) we present the ultraviolet luminosity functions (UVLFs) at high redshift ($z =7-20$) and compare them with observations, particularly those from \textit{JWST}. In Paper~III (Lu et al. in prep.) we present the redshift evolution of broadband LFs (from the FUV to the submillimetre) up to intermediate redshifts ($z \approx 6$).

This paper is organized as follows. In \autoref{sec:data_method}, we introduce the \colibre simulation suite (\autoref{sec:colibre}), the radiative transfer code, {\sc skirt}, which is used to model the interaction between starlight and dust in {\sc colibre} (\autoref{sec:skirt}), the galaxy sample selection method (\autoref{sec:sample}), the calculation of broadband and monochromatic luminosities (\autoref{sec:mono_lum}) and the compilation of observational data for comparison (\autoref{sec:obs_data}). The main results are presented in \autoref{sec:result}, where we study the correlation between galaxy stellar mass and multi-band luminosities (\autoref{sec:mstar2lum}) and the local LFs from the FUV to the $K$ band (\autoref{sec:fuv2k}), the mid-infrared (MIR) bands (\autoref{sec:mir}), the far-infrared (FIR) to submillimetre bands (\autoref{sec:fir2submm}) and the total infrared (TIR) LF (\autoref{sec:tir}). We discuss our results in \autoref{sec:discussion} and present a summary of our findings in \autoref{sec:conclusion}. Throughout this paper, we adopt the cosmological parameters used in the \colibre simulations (taken from \citealt{Abbott2022}): present-day matter density, $\Omega_{\rm m,0} = 0.306$; present-day baryon density, $\Omega_{\rm b,0} = 0.0486$; linear-theory amplitude of matter fluctuations on $8\,\mathrm{Mpc}/h$ scales, $\sigma_8 = 0.807$; and dimensionless Hubble parameter, $h = H_0/(100\,\textrm{km}\,\textrm{s}^{-1} \textrm{Mpc}^{-1}) =0.681$.

\section{Data and method}
\label{sec:data_method}
\subsection{The \colibre simulations}
\label{sec:colibre}
{\colibre} (COLd ISM and Better REsolution\footnote{\url{https://www.colibre-simulations.org/}}; \citealt{Schaye2026,Chaikin2026a}) is a new set of state-of-the-art cosmological hydrodynamical simulations of galaxy formation. These simulations are performed using the open-source code {\sc Swift} \citep{Schaller2024}, supplemented with the {\sc colibre} subgrid models for baryonic processes (gas cooling, star formation, chemical enrichment, stellar and AGN feedback, evolution of dust grains, etc.). The gas hydrodynamics equations are solved using the energy-density smoothed particle hydrodynamics (SPH) method {\sc Sphenix} \citep{Borrow2022}. The gravitational forces at short and long ranges are calculated differently: short-range forces are calculated with a 4$^{\rm th}$-order fast multipole method \citep{Cheng1999}, while long-range forces are calculated with a particle-mesh method in Fourier space. The initial conditions of \colibre were generated with the {\sc monofonIC} code \citep{Hahn2020,Michaux2021}, starting from redshift $z=63$, using second-order Lagrangian perturbation theory.

\subsubsection{Subgrid model}
\label{sec:subgrid}
Here we give a brief overview of the \colibre subgrid model. A full description can be found in \citet{Schaye2026}.

A key aspect of the \colibre simulations is that radiative cooling of gas is followed down to temperatures of about $10\,\mathrm{K}$ \citep{Ploeckinger2025}. This makes it possible to model explicitly the multiphase interstellar medium (ISM), including the cold dense component. Cooling and heating rates, together with the time evolution of chemical species, are calculated using the {\sc hybrid-chimes} implementation of the {\sc chimes} chemical network \citep{Richings2014a,Richings2014b}, adopting the ``weak interstellar radiation field (ISRF)'' model of \citet{Ploeckinger2025}. Hydrogen and helium are treated using non-equilibrium chemistry, while metal-line cooling is computed element by element assuming equilibrium ion fractions, with corrections for non-equilibrium free electron densities. The model includes molecule formation in the gas phase and on dust grains, shielding of radiation by gas and dust, photo-electric heating, photo-ionization and photo-dissociation, as well as heating from the metagalactic radiation field, the interstellar radiation field and cosmic rays. The local shielding length, radiation field strength and cosmic ray density are estimated from the local Jeans length, including both thermal and turbulent pressure support \citep{Schaye2001,Ploeckinger2020,Ploeckinger2025}.

Dust formation, growth and destruction are modelled self-consistently during the simulation using a dust grain evolution model that follows multiple grain sizes and species \citep{Trayford2026}. The dust model is coupled to the gas cooling calculations: chemical reactions and cooling processes involving dust depend on the local dust abundance and properties and metal-line cooling accounts for the depletion of individual elements onto dust grains.

Star formation is implemented using a Schmidt law together with a gravitational instability criterion \citep{Nobels2024}. Unlike the pressure-based star formation prescription of \citet{Schaye2008}, which is more appropriate when the cold ISM is not explicitly modelled (as in EAGLE and the majority of contemporary cosmological simulations of comparable volumes), this approach allows \colibre to predict an effective star formation law that emerges naturally from a resolved multiphase ISM \citep{Lagos2025}. As shown in \citet{Lagos2025}, \colibre reproduces the observed star formation laws well, both for galaxy-integrated measurements and for spatially resolved ones.

Stellar evolution and chemical enrichment are modelled using up-to-date nucleosynthetic yields from asymptotic giant branch stars, pre-supernova (SN) mass loss from massive stars, core-collapse SN and Type Ia SN \citep{Correa2026}. The delay time distribution of Type Ia SN is calibrated to reproduce the observed evolution of the cosmic SN Ia rate. Turbulent diffusion of both elemental abundances and dust mass fractions between neighbouring gas elements is also modelled.

Feedback from massive stars is included through several channels. Before and after the onset of core-collapse supernovae, early feedback from stellar winds, radiation pressure and \ion{H}{ii} regions is applied \citep{Benitez-Llambay2026}. Core-collapse supernova feedback includes a stochastic thermal component, based on \citet{DallaVecchia2012} and designed to reduce numerical overcooling but modified to better sample events in low-density gas, as well as a lower-energy kinetic component that provides additional turbulent support \citep{Chaikin2023}. Type Ia SN also inject energy through a stochastic thermal scheme to reduce numerical overcooling, although, unlike the core-collapse case where the energy per SN depends on the stellar birth gas pressure, the energy per Type Ia SN is fixed.

Black hole (BH) growth and AGN feedback are also included. Most of the \colibre simulations use a purely thermal AGN feedback model \citep{Booth2009}, modified to improve feedback sampling for low-mass black holes. A subset of simulations adopt a hybrid AGN feedback model that combines thermal feedback (representing the effects of AGN winds) with kinetic jets \citep{Husko2026}. In this model, black hole spin evolution is tracked and different BH accretion disk states are treated separately. In this work, we use the \colibre simulations with the purely thermal AGN feedback model as the default. We confirm that the different AGN feedback models have little influence on the LFs across all bands (see a comparison of LFs based on different AGN feedback models in Appendix~\ref{sec:agn}).

\subsubsection{Calibration}

The calibration of the subgrid supernova and AGN feedback model in \colibre is described in detail in \citet{Chaikin2026a}. Here, we briefly describe the calibration strategy. The {\sc colibre} model is calibrated by adjusting up to four subgrid parameters (see Table~1 of \citealt{Chaikin2026a} for the list of parameters) that control the strengths of stellar and AGN feedback in order to reproduce simple observed galaxy scaling relations at $z=0$: the galaxy stellar mass function (GSMF) and the galaxy size-stellar mass relation (SSMR). The observed GSMF is taken from \citet{Driver2022}, while the SSMR is constrained from \citet{Hardwick2022}. The calibration is first carried out at the lowest resolution level of {\sc colibre}, called m7, which has $m_{\rm g} = 1.47 \times 10^{7}\,\mathrm{M_{\odot}}$ and $m_{\rm DM} = 1.94 \times 10^{7}\,\mathrm{M_{\odot}}$ (see \citealt{Schaye2026} for more details of the \colibre data products). Following \citet{Kugel2023}, Gaussian process emulators are first trained on the m7 simulations that sample the parameter space via a Latin hypercube design, enabling the emulator to predict the GSMF and SSMR as continuous functions of the free parameters. The best-fitting parameters are then obtained by minimizing the difference between the emulator predictions and the observational constraints within the stellar mass range $10^{9}$ to $10^{11.3}\,\rm M_{\odot}$.

After calibrating the m7 model, higher-resolution simulations are not re-calibrated using emulators. Instead, the best-fitting m7 parameters are adopted as reference points and the SN and AGN feedback parameters are adjusted slightly by hand to recover a comparable level of agreement with the $z = 0$ GSMF and SSMR across different resolutions. Finally, the model is calibrated by hand to match the local observed black hole mass-stellar mass relation of \citet{Graham2023}, primarily by setting the AGN feedback coupling efficiency; the black hole seed mass also affects the predicted black hole mass-stellar mass relation.

\subsubsection{Structure finding}
To find galaxies in the \colibre simulations, a friends-of-friends halo finder \citep{Davis1985} is first applied to locate dark matter halos. Self-bound structures (subhalos) are then identified using the subhalo finder {\sc hbt-herons} \citep{Han2018,Forouhar_Moreno2025}. Unlike methods that use only position or phase-space information from a single snapshot, {\sc hbt-herons} tracks subhalos over time by following their particles. Readers are referred to \citet{Schaye2026} (Section 2.3) for more details of structure finding in {\sc colibre}. Throughout this work, we define a galaxy as the set of baryonic and dark matter particles belonging to one such self-bound subhalo.

\subsubsection{Data products}
\label{sec:colibre_data_product}
The \colibre simulations have been performed at three resolutions: m5 (highest resolution; gas and dark matter particle masses $m_{\rm g} = 2.3 \times 10^{5}\,\mathrm{M_{\odot}}$ and $m_{\rm DM} = 3 \times 10^{5}\,\mathrm{M_{\odot}}$); m6 (medium resolution; $m_{\rm g} = 1.8 \times 10^{6}\,\mathrm{M_{\odot}}$ and $m_{\rm DM} = 2.4 \times 10^{6}\,\mathrm{M_{\odot}}$); and m7 (lowest resolution; $m_{\rm g} = 1.47 \times 10^{7}\,\mathrm{M_{\odot}}$ and $m_{\rm DM} = 1.94 \times 10^{7}\,\mathrm{M_{\odot}}$) and are available in different cube sizes (25, 50, 100, 200 and 400 $\rm cMpc$), referred to as L025, L050, L100, L200 and L400, respectively (see Table 2 of \citealt{Schaye2026} for full details of all \colibre simulations). For each simulation, dark matter is supersampled compared to gas particles by a factor of 4, yielding similar dark matter and baryonic particle masses, in order to suppress spurious transfer of energy from dark matter to stars \citep{Ludlow2019,Ludlow2023}.

\subsection{The radiative transfer code: {\sc skirt}}
\label{sec:skirt}
\begin{figure*}
\centering
\includegraphics[width=0.82\textwidth]{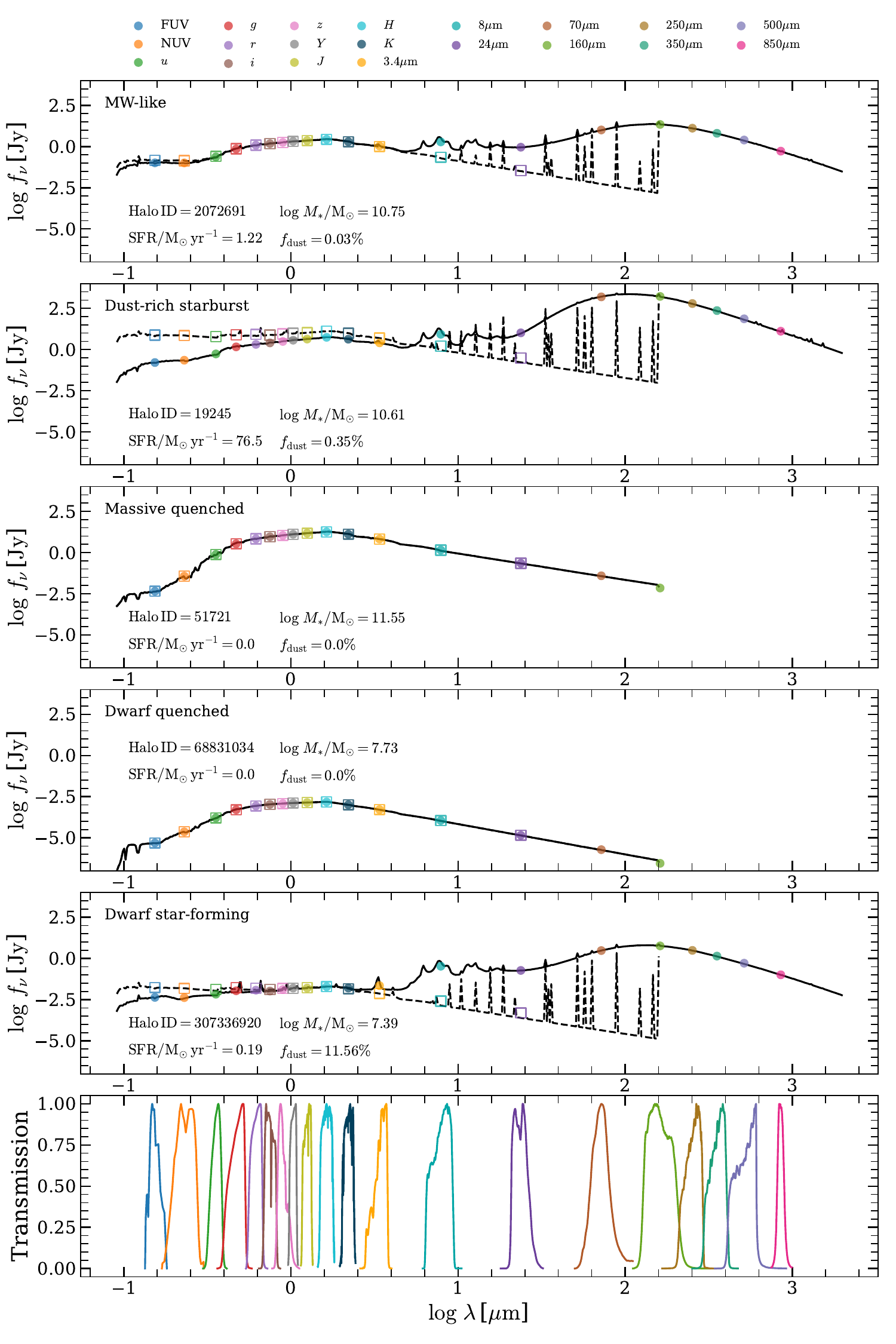}\vspace{-0.2cm}
\caption{Example SEDs produced with \skirt for five galaxies from L200m6 (from top to bottom: a Milky Way-like star-forming galaxy, a dust-rich starburst galaxy, a massive quenched galaxy, a quenched dwarf galaxy and a dust-rich star-forming dwarf galaxy). The halo ID, stellar mass, SFR and the dust-to-stellar mass ratio ($f_{\rm dust}$) of each galaxy are shown in each panel. The galaxies are ``observed'' with a mock detector placed at a distance of 10 Mpc from the galaxy centre. For each galaxy, we show the SEDs with (solid) and without (dashed) dust. Coloured points indicate the broadband luminosities in the different filters (filled circles show results that include dust, while open squares show dust-free results; for the dust-free case, we only show bands with pivot wavelengths up to $24\,\micron$). The transmission curves of all filters are shown in the bottom panel. The filters and their pivot wavelengths are listed in \autoref{table:filter}.}
\label{fig:example_sed}
\end{figure*}

In this study, we use the {\sc colibre-skirt} pipeline described in \citet{Gebek2026} to model the interaction between stellar radiation and dust in {\sc colibre} galaxies. This pipeline is built upon the \skirt code\footnote{Available at \url{https://skirt.ugent.be}.} (version 9; \citealt{Camps2020}), which is a publicly available, flexible 3D Monte Carlo radiative transfer code designed to model the interaction between radiation and matter in astrophysical systems. The pipeline accounts for dust absorption, scattering, heating and thermal re-emission, based on the dust distributions predicted self-consistently by the \colibre simulations. Unlike previous studies in which \skirt was applied to earlier simulations (e.g. \citealt{Camps2016,Camps2018,Trcka2022}), the {\sc colibre-skirt} pipeline is calibration-free, in the sense that no free parameters in the {\sc colibre-skirt} post-processing are tuned to reproduce the LFs studied here. Hence, any agreement with the observed LFs is not enforced by construction, but instead reflects the predictive power of the model. Below we briefly summarize the main aspects of the {\sc colibre-skirt} pipeline. We refer readers to \citet{Gebek2026} for full details.

We decompose the luminosity sources in galaxies into three components: (1) evolved stars, (2) star-forming regions and (3) dust. The evolved stars are defined to be the stellar particles older than a threshold age of $t_{\rm R}=10\,\mathrm{Myr}$. Their emission is modelled by assigning a simple stellar population (SSP) template from BPASS (v2.2.1; \citealt{Eldridge2017,Stanway2018}) to each stellar particle, consistently with the early stellar feedback module adopted in \colibre \citep{Benitez-Llambay2026} and assuming a \citet{Chabrier2003} IMF, as in {\sc colibre}. 

Stellar populations younger than $t_{\rm R}$ are resampled using the star-forming gas to better capture the emission from young stars, which are poorly sampled by the stellar particles at the mass resolution of {\sc colibre}. The emission from these star-forming regions is computed using the {\sc toddlers v1} library \citep{Kapoor2023,Kapoor2024}, which models the time-dependent evolution of \ion{H}{ii} regions around young stars and includes the emission lines from the ionised gas in the \ion{H}{ii} regions. The {\sc toddlers} library is also based on the BPASS (v2.2.1) library, consistent with what we use for evolved stars. In this work, we adopt the dust-free version of {\sc toddlers} (which currently does not include nebular continuum emission\footnote{A new version (Version 2) of the {\sc toddlers} library is now available \citep{Kapoor2026}, with nebular continuum emission included in the v2-DTM dust-free model (see their Table E.2). We plan to adopt this updated model in future applications.}; see \citealt{Gebek2026}), with a cloud density of $n_{\rm cl}=320~\mathrm{cm^{-3}}$ and a star-formation efficiency of $\eta_{\rm sf}=2.5\%$, such that attenuation and emission by the dust in {\sc colibre} are treated self-consistently in the subsequent {\sc skirt} radiative transfer calculation. The luminosities of star-forming regions are determined using the SFR averaged over the preceding time interval $t_{\rm R}$, assuming a constant star formation history over this period. We find that the star-forming region resampling results in a $\approx 0.4 \,\rm dex$ offset in luminosity in the FUV band at $z=0$, with the LFs in the other bands remaining essentially unchanged (see Appendix~\ref{sec:resampling} for a comparison of the LFs with and without star-forming region resampling).

\colibre tracks dust of three compositions (graphite, forsterite, fayalite, corresponding to carbonaceous, magnesium-silicate and iron-silicate, respectively) at two grain sizes ($0.01\,\mu\mathrm{m}$ and $0.1\,\mu\mathrm{m}$), yielding six dust grain types. \skirt uses the position of each gas particle and four different dust species masses (combining the \colibre Mg- and Fe-rich silicates into a single silicate species). The optical properties and grain size distributions of the dust grains are based on \citet{Draine2007a}. These size distributions are adjusted for each individual gas particle by splitting the original \citet{Draine2007a} distribution at $a=0.03\,\mu\mathrm{m}$ (where $a$ is the dust grain size), separating the silicate and graphite grains into small and large components (``split \& scale''; see \citealt{Gebek2026} for more details). Each of these four components is then assigned the exact dust mass predicted by \colibre for the corresponding gas particle. The dust temperature, on which the dust emission depends, is computed self-consistently by \skirt from the local radiation field. In the fiducial \cosk pipeline, however, secondary dust heating through the re-absorption of dust-emitted photons is disabled. We find that this approximation has a negligible effect on the MIR luminosities, but produces a modest increase in the FIR and total infrared luminosities of the brightest galaxies (see \autoref{sec:mismatch_mir}).

Emission from polycyclic aromatic hydrocarbons (PAHs), arising from very small carbonaceous grains that are stochastically heated, is an important contributor to the MIR emission and therefore likely plays a key role in shaping the MIR LFs. Modelling this emission is challenging, however, because PAHs are not explicitly tracked in {\sc colibre} and the MIR emission is sensitive to both the abundance of these very small grains and their detailed optical and calorimetric properties. In our fiducial pipeline, PAH emission is also modelled using the \citet{Draine2007a} dust model. We assign $\approx 60\%$ of the small carbonaceous dust predicted by \colibre to a PAH component\footnote{This fraction is not treated as a free parameter in our pipeline, but is set by the PAH mass fraction within the small carbonaceous grain population in the \citet{Draine2007a} dust model.}, split equally between neutral and ionised PAHs. The resulting PAH emission is then calculated self-consistently by {\sc skirt}, including stochastic heating.

The dust-processed emission of galaxies in \colibre is recorded by an ideal mock detector placed along the $Z$-axis of the simulation cube, corresponding to a random galaxy orientation, at a distance of $10$ Mpc. We do not include PSF convolution or noise injection in this analysis. In this work, we compute galaxy luminosities within a fixed projected circular aperture of proper radius $50$~kpc. We note that only gravitationally bound particles are included in our analysis. We find that the LFs, especially at the bright end, from the FUV to the $K$ band are influenced by the choice of aperture size, while those in the IR bands are not. A detailed analysis of the impact of varying the aperture size on the LFs is provided in Appendix~\ref{sec:aperture}.

In \autoref{fig:example_sed}, we present the SEDs (from \skirt modelling; calculated within a projected aperture of radius 50 kpc) of five example galaxies from the \colibre L200m6 simulation: a star-forming galaxy with a Milky Way-like stellar mass and SFR, a dust-rich starburst galaxy, a massive quenched galaxy, a dust-poor quenched dwarf galaxy and a dust-rich star-forming dwarf galaxy. For each galaxy, we show the SEDs both with (solid) and without (dashed) dust, along with the filter-integrated fluxes centred at the pivot wavelengths (see \autoref{sec:mono_lum} for details). Since dust emission is dominant at longer wavelengths, we only present the filter-integrated fluxes for SEDs (and LFs in later sections) {\it without} dust for wavelengths up to $24\,\micron$. As the wavelength range covered by the BPASS models used to model evolved stars extends only to $\lambda = 160\,\micron$, we truncate the dust-free SEDs at 160\,$\micron$. The emission lines seen in the dust-free SEDs originate from \ion{H}{ii} regions, as well as from the surrounding neutral gas in star-forming regions heated by non-ionizing radiation from young stars and are predicted by the {\sc toddlers} models.

As can be seen, galaxies with distinct star formation histories and dust content exhibit remarkably different SEDs. The MW-like galaxy (first panel) has clear PAH emission in the wavelength range $\approx 3-20\,\micron$ and significant emission at both short wavelengths (e.g. UV bands) and long wavelengths (e.g. FIR bands); a moderate dust attenuation effect is also visible in the UV. The dust-rich starburst galaxy (second panel) shows significantly stronger IR emission and greater UV attenuation due to its higher dust content. Furthermore, its UV emission is notably stronger than that of the MW-like galaxy because of its much higher SFR. The enhanced FIR emission results from dust heating through the absorption of intense UV radiation from young stars. The quenched massive galaxy (third panel) shows clear emission in the optical bands but weak emission in the UV and IR bands, consistent with its low star formation and dust content. The dust-poor quenched dwarf galaxy (fourth panel) shows a similar SED shape to that of the massive quenched galaxy, but with a significantly lower amplitude due to its lower stellar mass. Similarly, the star-forming dust-rich dwarf galaxy (fifth panel) shows a similar SED shape to the dust-rich starburst galaxy (strong IR emission and UV dust attenuation), but also with an overall lower amplitude. All these spectral characteristics are consistent with theoretical expectations for these various galaxy types, reflecting the realism of the \skirt modelling.

\begin{figure*}
\centering
\includegraphics[width=0.8\textwidth]{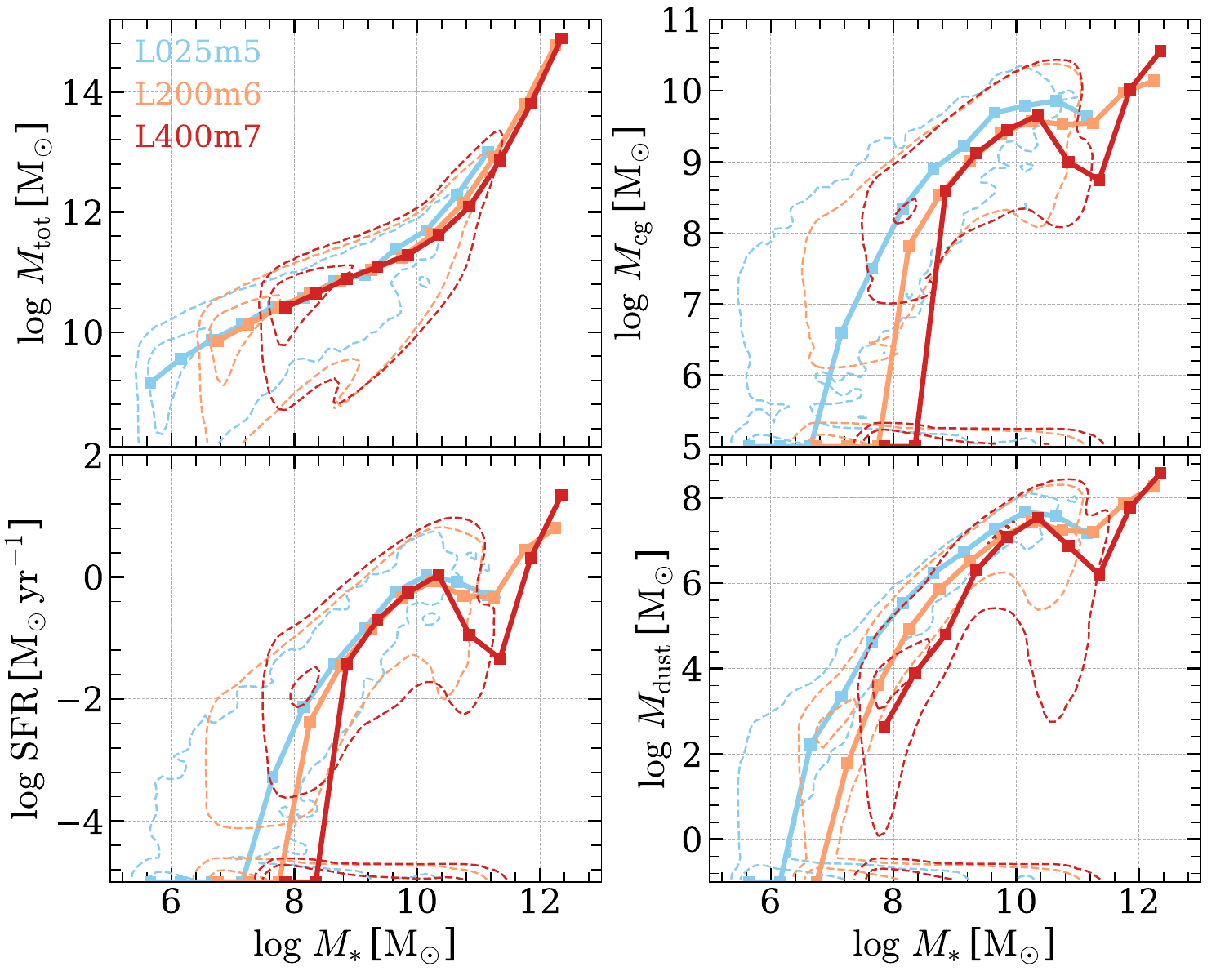}\vspace{-0cm}
\caption{Relationships between galaxy stellar mass and other global properties: (1) total mass (including stars, gas, dark matter and black holes), $M_{\rm tot}$ (top left); (2) cool dense gas mass ($T < 10^{4.5}\,\mathrm{K}$ and $\rho_{\rm g}/m_{\mathrm{H}} > 10^{-1}\,\mathrm{cm}^{-3}$), $M_{\rm cg}$ (top right); (3) instantaneous SFR (bottom left); and (4) total dust mass, $M_{\rm dust}$ (bottom right) of \colibre galaxies (above the stellar-mass limits listed in \autoref{table:simulations}) at $z=0$. The total mass is calculated using all gravitationally bound particles, while the other quantities are measured within a spherical aperture of radius 50 kpc. Data points show medians in each stellar-mass bin and the underlying contours show the distributions of all galaxies presented. Good convergence is seen in the total mass-stellar mass relation and, at intermediate stellar masses, in the SFR - stellar mass relation, whereas the cool gas mass - stellar mass and dust mass - stellar mass relations show a clear resolution dependence.}
\label{fig:Mstar2Pros}
\end{figure*}

\begin{figure*}
\centering
\includegraphics[width=\textwidth]{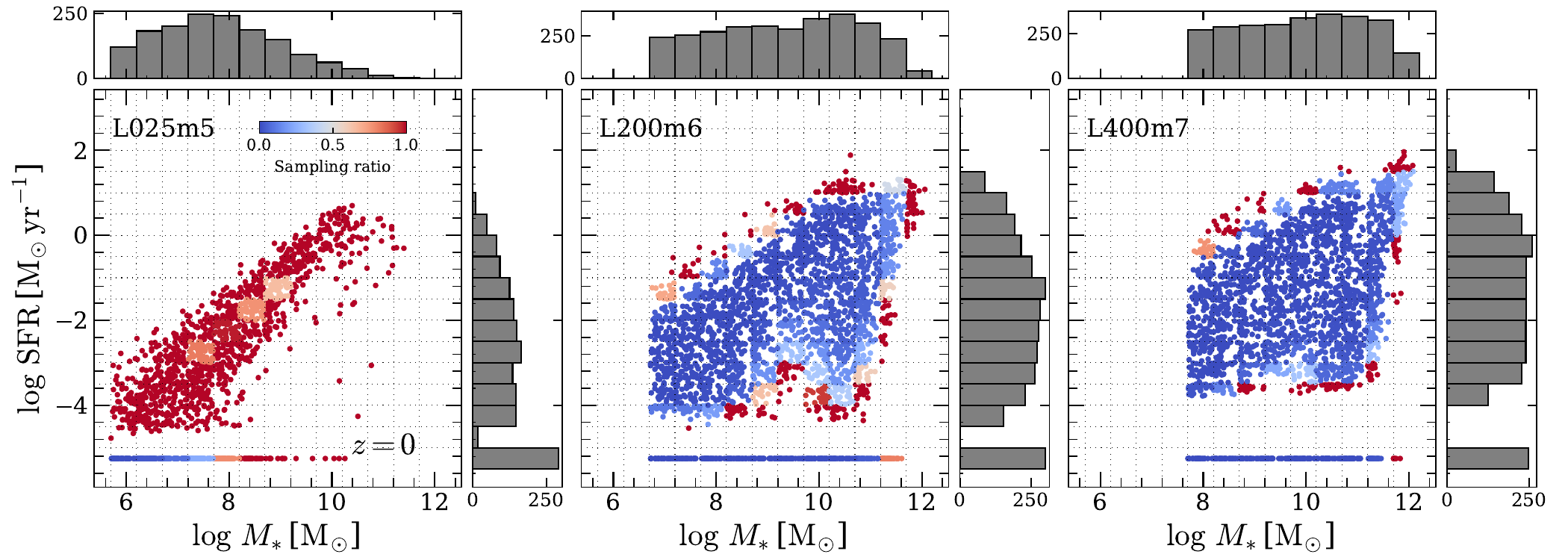}\vspace{-0.6cm}
\caption{The distribution of galaxies selected for post-processing by \skirt on the $\log\,M_{\ast}-\log \,\rm SFR$ plane for L025m5, L200m6 and L400m7 \colibre simulations (from left to right) at $z = 0$. For visualization purposes, SFRs lower than $10^{-5.25}\,\mathrm{M_{\odot}\,yr^{-1}}$ are set to $10^{-5.25}\,\mathrm{M_{\odot}\,yr^{-1}}$. The dashed lines indicate the boundaries of the stellar mass and SFR bins used in the sample selection. The colours represent the sampling fraction of galaxies (the ratio between the number of galaxies selected from the bin and the total number of galaxies in the bin). The histograms show the distributions of stellar mass and SFR for the selected galaxies.}
\label{fig:sample}
\end{figure*}
\subsection{Sample selection}
\label{sec:sample}

To carry out the {\sc skirt} radiative transfer calculation, we select a sample of galaxies from three \colibre simulations of different volume and mass resolution at $z=0$ (L025m5, L200m6 and L400m7)\footnote{L200 and L400 are the largest available volumes at resolution m6 and m7, respectively. For m5, the largest and second-largest volumes (L100 and L050) are still running at the time of writing and have not yet reached $z=0$. Therefore, in this work we analyse L025m5. The \colibre galaxy data were accessed using the {\sc python} packages {\sc swiftsimio} \citep{Borrow2020,Borrow2021} and {\sc swiftgalaxy} \citep{Oman2025}.}, as listed in \autoref{table:simulations}.

In \autoref{fig:Mstar2Pros}, we present correlations between galaxy stellar mass and other global galaxy properties in simulations at different resolutions (for galaxies of stellar mass above $5\times 10^{5}\,\rm M_{\odot}$ for m5, $5\times 10^{6}\,\rm M_{\odot}$ for m6 and $5\times 10^{7}\,\rm M_{\odot}$ for m7, corresponding roughly to $2-6$ stellar particles\footnote{The approximate number of stellar particles in galaxies at the lower stellar-mass boundary varies slightly with resolution.}). These include (1) the total mass (including stars, gas, dark matter and black holes), $M_{\rm tot}$; (2) the mass of gas in the cool, dense phase ($T < 10^{4.5}\,\mathrm{K}$ and $\rho_{\rm g}/m_{\mathrm{H}} > 10^{-1}\,\mathrm{cm}^{-3}$, where $T$ and $\rho_{\rm g}$ are the temperature and density respectively and $m_{\mathrm{H}}$ is the mass of a hydrogen atom), $M_{\rm cg}$; (3) the instantaneous SFR; and (4) the dust mass, $M_{\rm dust}$. $M_{\rm tot}$ is calculated using all the gravitationally bound particles of the galaxy, while the other global properties are calculated within a sphere of radius 50 kpc. All quantities are taken from the Spherical Overdensity and Aperture Processor (SOAP; \citealt{McGibbon2025}) catalogues. 

As can be seen, the three \colibre simulations at different resolutions show remarkable consistency in the total mass - stellar mass relation across the available stellar mass range and overall good convergence in the relation between SFR and stellar mass at intermediate stellar masses. The disagreement at the low stellar mass end is likely due to the incompleteness of the galaxy sample caused by the resolution limit. At the high-mass end, m5 and m6 still show good agreement, while m7 deviates from the other two. This deviation is likely related to AGN feedback, which is driven by dense gas in the galaxy nucleus and is therefore sensitive to numerical resolution. The convergence improves again at the highest stellar masses ($M_{\ast} \sim 10^{12}\,\mathrm{M_\odot}$), where AGN activity is less important. For cool gas mass and dust mass, the relations show a clear resolution dependence. This may be because higher numerical resolution (lower particle mass) better captures the formation of cool and dense ISM phases (that lower resolution counterparts do not), where dust growth by accretion is efficient. As a result, at fixed stellar mass, the higher-resolution simulations can build up larger cool gas reservoirs and higher dust masses, leading to the observed resolution dependence. We confirm that the results presented here are not driven by the different cube sizes of simulations at different resolutions: similar trends are also seen in comparisons among the m5, m6 and m7 simulations with the same cube size (L025).

Running radiative transfer modelling for all galaxies in the whole simulation volume can be time-consuming and computationally expensive (see \citealt{Gebek2026} for an estimate), while LFs can be accurately calculated with only a representative subsample. Thus, for each \colibre simulation, we select a subsample of galaxies (regardless of whether they are centrals or satellites) following the steps below:

\begin{enumerate}
\item We first exclude galaxies with stellar masses below the same stellar mass thresholds given above for each \colibre simulation ($5\times 10^{5}\,\rm M_{\odot}$ for m5, $5\times 10^{6}\,\rm M_{\odot}$ for m6 and $5\times 10^{7}\,\rm M_{\odot}$ for m7).
\item The remaining galaxies are then binned by their stellar mass and instantaneous SFR, with a bin size of 0.5 dex for both stellar mass and SFR. From each bin in the stellar mass-SFR grid, we randomly select a specific target number (50 for m5 and 30 for m6 and m7) of galaxies out of the whole galaxy population. If a bin contains fewer galaxies than the target number, all available galaxies are included.
\item For each stellar mass-SFR bin, we calculate the sampling fraction as the ratio between the number of selected galaxies and the total number of galaxies in this bin. When computing the LFs, the inverses of these sampling fractions are used to assign a weight to each selected galaxy contribution, thereby preserving the overall galaxy population statistics (see \autoref{eq:lf}). The total number of selected galaxies and their {\it average} fractions relative to the {\em total} galaxy population in the different \colibre simulations are listed in \autoref{table:simulations}.
\end{enumerate}

We note that galaxies are binned in the stellar mass-SFR plane, rather than by stellar mass alone, to ensure coverage of systems at different stages of star formation activity within each mass range. This prevents the sample from being dominated by either actively star-forming or quiescent galaxies and enables a more representative sampling of the stellar and dust emission that contributes to the LFs from the FUV to the submillimetre. The distributions of the selected galaxy samples from the three \colibre simulations in the $\log\,M_{\ast}-\log\,\rm SFR$ plane are shown in \autoref{fig:sample}, with the colours indicating the sampling fractions. The target number (50 for m5 and 30 for m6 and m7) of selected galaxies in each stellar mass-SFR bin used here is chosen to keep the computational cost low while still providing sufficient sampling to allow an accurate calculation of the LFs. We have verified that the selected sample is sufficient for an accurate estimate of the LFs of the simulation cube across the whole wavelength range (see Appendix~\ref{sec:random_sampling} for a test). 

\begin{table*}
\caption{The \colibre simulations used in this work. Column (1): simulation name; column (2): size of the cosmological cube per dimension; column (3): mean initial gas particle mass; column (4): mean DM particle mass; column (5): initial number of gas particles; column (6): number of DM particles; column (7): AGN feedback model used in the simulation; column (8): stellar mass limit for selecting galaxies; column (9): the number of selected galaxies and the {\it average} sampling ratio at $z=0$.} \setlength{\tabcolsep}{2mm}
\begin{tabular}{lrccrrccc}
\hline
\multicolumn{1}{c}{Simu. Name} &
\multicolumn{1}{c}{$L_{\rm cube} [\mathrm{cMpc}]$} &
\multicolumn{1}{c}{$m_{\rm gas} [\mathrm{M_{\odot}}]$} &
\multicolumn{1}{c}{$m_{\rm DM} [\mathrm{M_{\odot}}]$} &
\multicolumn{1}{c}{$N_{\rm gas}$} &
\multicolumn{1}{c}{$N_{\rm DM}$} &
\multicolumn{1}{c}{AGN feedback model} &
\multicolumn{1}{c}{$M_{\ast,\rm min}[\mathrm{M_{\odot}}]$} &
\multicolumn{1}{c}{$N_{\rm selected}$ (sampling ratio)}\\
\hline
L025m5 & 25 & $2.30\times 10^5$  & $3.03\times 10^5$ & $752^3$ & $4\times 752^3$ & Thermal & $5\times 10^{5}$ & 1540 (17.79\%)\\
\hline
L100m6 & 100 & $1.84\times 10^6$  & $2.42\times 10^6$ & $1504^3$ & $4\times 1504^3$ & Thermal & $5\times 10^{6}$ & 2399 (1.40\%)\\
\hline
L200m6 & 200 & $1.84\times 10^6$  & $2.42\times 10^6$ & $3008^3$ & $4\times 3008^3$ & Thermal & $5\times 10^{6}$ & 2994 (0.22\%)\\
\hline
L400m7 & 400 & $1.47\times 10^7$  & $1.94\times 10^7$ & $3008^3$ & $4\times 3008^3$ & Thermal & $5\times 10^{7}$ & 2654 (0.07\%)\\
\hline
L100m6h & 100 & $1.84\times 10^6$  & $2.42\times 10^6$ & $1504^3$ & $4\times 1504^3$ & Hybrid & $5\times 10^{6}$ & 2512 (1.48\%)\\
\hline
\end{tabular}
\label{table:simulations}
\vspace{2mm}
\end{table*}

\subsection{Calculation of monochromatic and broadband luminosities}
\label{sec:mono_lum}
In observational studies, luminosities used to derive LFs are typically obtained from broadband photometry, but their definitions vary across wavelength ranges. From the FUV to the $K$ band (e.g. \citealt{Budavari2005,Driver2012}) and in some IR studies (e.g. \citealt{Dai2009,Moore2026}), the quoted luminosities are usually filter-averaged quantities, that is, broadband luminosities associated with the corresponding broadband filters. By contrast, some studies in the (mid- and far-)IR bands (e.g. \citealt{Marleau2007,Negrello2013}) adopt so-called monochromatic luminosities, which are defined at a specific wavelength or frequency and derived from the broadband measurements. 

One commonly adopted approach to derive monochromatic luminosities is to assume a specific SED shape, typically a power law with a fixed spectral index, $L_{\nu} = A \nu^{\alpha}$, where $L_{\nu}$ is the luminosity density at frequency $\nu$ and $A$ is the normalization factor. For example, \citet{Negrello2013} assume $\alpha = 3$ for submillimetre galaxies, while \citet{Marleau2007} adopt $\alpha = -0.7$ for mid-infrared wavelengths. The normalization factor $A$ can then be determined by requiring that the filter-weighted integral of the assumed SED reproduces the observed broadband flux, after which the monochromatic luminosity at a target frequency (or wavelength) is obtained by evaluating the normalized SED at that frequency (or wavelength). This approach is expected to work better at longer wavelengths, such as in the submillimetre regime where SEDs are relatively smooth, whereas in the mid-infrared the complex spectral structure introduced by PAH emission can make the assumed SED shape less representative and thus introduce additional uncertainties (e.g. \citealt{Draine2007a,Draine2007b,Smith2012}).

An alternative approach is to directly read off the monochromatic luminosities from SEDs that are fitted to broadband luminosities (by using the broadband filters to compute the broadband luminosities from the SEDs; e.g. \citealt{Rodighiero2010,Marchetti2016}). The main drawback of this approach is that the derived monochromatic luminosities are not direct observables but can depend sensitively on the assumed or fitted SED, with the uncertainties being particularly significant in wavelength regimes where the SED exhibits strong spectral features, such as in the mid-infrared.

In this work, we follow the practice of \citet{Camps2016} (see also \citealt{Trcka2020,Gebek2024}) and adopt filter-weighted mean luminosities across all bands from the FUV to the submillimetre. For the IR and submillimetre bands, we use these filter-weighted mean luminosities based on filters with pivot wavelengths (see \autoref{eq:pivot} for the definition) close to the target wavelengths, in order to approximate the corresponding monochromatic luminosities. While this quantity represents an average over the finite filter bandwidth rather than a strictly monochromatic value, it is well defined, does not rely on any assumption about the SED shape and is directly comparable to observational measurements. We investigate the influence on LFs of using different monochromatic luminosity definitions in Appendix~\ref{sec:mono_lum_appendix} and confirm that different monochromatic luminosity definitions only have a minor influence on the LFs. 

We note that, as pointed out on Ivan K. Baldry’s web page\footnote{\url{https://www.astro.ljmu.ac.uk/~ikb/research/mags-fluxes/}} and by \citet{Camps2016} (see their Appendix~A; see also \citealt{Tokunaga2005}), obtaining the band-integrated flux from the SED and the instrument response curve requires different formulae depending on whether the instrument is a photon counter or a bolometer. In this work, we first transform all response curves of filters used with photon-counting instruments into the corresponding bolometer response curves (see \citealt{Camps2016}) using:
\begin{equation}
T(\lambda) = \lambda R(\lambda),
\end{equation}
where $T(\lambda)$ and $R(\lambda)$ denote the filter throughputs of the bolometers and photon-counting filters, respectively, as functions of wavelength. The band-averaged specific flux is then calculated as:
\begin{equation}
\langle F_\lambda \rangle = \frac{\int F_\lambda(\lambda)\, T(\lambda)\, \mathrm{d}\lambda}{\int T(\lambda)\, \mathrm{d}\lambda},
\end{equation}
where $F_\lambda(\lambda)$ is the specific flux at wavelength $\lambda$. Note that the signal measured by a detector is directly proportional to $\int F_\lambda(\lambda) T(\lambda) \mathrm{d}\lambda$. Finally, $\langle F_\lambda \rangle$ is converted to $\langle F_\nu \rangle$ using
\begin{equation}
\langle F_\nu \rangle = \langle F_\lambda \rangle \frac{\lambda_{\rm pivot}^2}{c},
\end{equation}
where $c$ is the speed of light and $\lambda_{\rm pivot}$ is the pivot wavelength, defined as:
\begin{equation}
\label{eq:pivot}
\lambda_{\text {pivot }}=\sqrt{\frac{\int T(\lambda) \mathrm{d} \lambda}{\int T(\lambda) \mathrm{d} \lambda / \lambda^2}}.
\end{equation}

In \autoref{table:filter}, we list all the filters used to compute the monochromatic LFs in this work, together with their pivot wavelengths and bandwidths. We note that, for a given wavelength, there can be multiple available filters (for example, {\sc spitzer\_mips\_70} and {\sc herschel\_pacs\_70} at $\lambda_{\rm pivot} \approx 70\,\mathrm{\mu m}$). We do not have a preference for any particular choice; however, we have verified that using different filters with the same pivot wavelength has little impact on the derived LFs. The bandwidths quoted in \autoref{table:filter} are the passband rectangular widths, calculated following the HST WFC3 Instrument Handbook\footnote{See the WFC3 Instrument Handbook, Section~6.5
(\url{https://hst-docs.stsci.edu/wfc3ihb/chapter-6-uvis-imaging-with-wfc3/6-5-uvis-spectral-elements}).} as
\begin{equation}
\label{eq:band_width}
\Delta \lambda = \frac{\int T(\lambda)\,\mathrm{d}\lambda}{\max[T(\lambda)]}.
\end{equation}
 
\begin{table*}
\caption{Summary of the bands and observed LF data used in this work. Column~(1): band short name; (2): adopted filter; (3): pivot wavelength (defined in \autoref{eq:pivot}); (4): bandwidth (defined in \autoref{eq:band_width}); and (5): references for the corresponding observational datasets and their redshift ranges. The filters listed here are adopted for monochromatic luminosity calculations and are not necessarily the same as those used in the observational studies. We have verified that the choice of filters at a given pivot wavelength does not affect the LFs.}\setlength{\tabcolsep}{2.5mm}
\begin{tabular}{ccccp{6cm}}
\hline
\hline
 Name & Filter & $\lambda_{\rm pivot}\,[\mathrm{\mu m}]$ & $\Delta\lambda\,[\mathrm{\mu m}]$ & References [redshift range]\\
\hline
FUV & GALEX FUV &  0.15351 &  0.02656 & \citet{Wyder2005} [0,0.1], \citet{Budavari2005} [0.07,0.13], \citet{Driver2012} [0.013,0.1]\\
\hline
NUV & GALEX NUV &  0.23008 &  0.07683 & \citet{Wyder2005} [0,0.1], \citet{Budavari2005} [0.07,0.13], \citet{Driver2012} [0.013,0.1]\\
\hline
$u$ & SLOAN SDSS $u$ &  0.35565&  0.05410 & \citet{Driver2012} [0.013,0.1], \citet{Loveday2012} [$0.002-0.1$]\\
\hline
$g$ & SLOAN SDSS $g$ &  0.47024&  0.10646 & \citet{Driver2012} [0.013,0.1], \citet{Loveday2012} [$0.002-0.1$], \citet{Moore2026} [corrected to $z=0.1$]\\
\hline
$r$ & SLOAN SDSS $r$ &  0.61755&  0.10555 & \citet{Driver2012} [0.013,0.1], \citet{Loveday2012} [$0.002-0.1$], \citet{Moore2026} [corrected to $z=0.1$]\\
\hline
$i$ & SLOAN SDSS $i$ &  0.74899&  0.11025 & \citet{Driver2012} [0.013,0.1], \citet{Loveday2012} [$0.002-0.1$]\\
\hline
$z$ & SLOAN SDSS $z$ &  0.89467&  0.11640 & \citet{Driver2012} [0.013,0.1], \citet{Loveday2012} [$0.002-0.1$], \citet{Moore2026} [corrected to $z=0.1$]\\
\hline
$Y$ & UKIRT UKIDSS Y &  1.0314&  0.0970 & \citet{Hill2010} [$<0.1$], \citet{Driver2012} [0.013,0.1]\\
\hline
$J$ & UKIRT UKIDSS J &  1.2501&  0.1432& \citet{Hill2010} [$<0.1$], \citet{Driver2012} [0.013,0.1], \citet{Jones2006} [0.0025,0.2]\\
\hline
$H$ & UKIRT UKIDSS H &  1.6354&  0.2772& \citet{Hill2010} [$<0.1$], \citet{Driver2012} [0.013,0.1], \citet{Jones2006} [0.0025,0.2]\\
\hline
$K$ & UKIRT UKIDSS K &  2.2058&  0.3179& \citet{Hill2010} [$<0.1$], \citet{Driver2012} [0.013,0.1], \citet{Jones2006} [0.0025,0.2]\\
\hline
$3.4\mathrm{\mu m}$ & WISE W1 &  3.3897&  0.6215 & \citet{Moore2026} [corrected to $z=0.1$], \citet{Babbedge2006} (3.6 \micron) [0,0.25], \citet{Dai2009} (3.6 \micron) [0,0.2]\\
\hline
$8\mathrm{\mu m}$ & SPITZER IRAC I4 &  7.8842&  2.3454 & \citet{Babbedge2006} [0,0.25], \citet{Dai2009} [0,0.2], \citet{Rodighiero2010} [0,0.3]\\
\hline
$15\mathrm{\mu m}$ & ISO LW3 &  14.404&  4.621 & \citet{Pozzi2004} [corrected to $z=0$], \citet{Rodighiero2010} [0,0.3], \citet{Wu2011} [0.02, 0.3]\\
\hline
$24\mathrm{\mu m}$ & SPITZER MIPS 24 &  23.759&  5.581 & \citet{Rodighiero2010} [0,0.25], \citet{Babbedge2006} [0,0.25], \citet{Marchetti2016} [0.02,0.1], \citet{Wu2011} [0.02, 0.3]\\
\hline
$70\mathrm{\mu m}$ & SPITZER MIPS 70 &  71.987&  21.5 & \citet{Patel2013} [0,0.2], \citet{Marchetti2016} [0.02,0.2]\\
\hline
$160\mathrm{\mu m}$ & HERSCHEL PACS 160 &  161.89&  69.76 & \citet{Patel2013} [0,0.2], \citet{Marchetti2016} [0.02,0.2]\\
\hline
$250\mathrm{\mu m}$ & HERSCHEL SPIRE 250 &  252.55&  60.65 & \citet{Vaccari2010} [0,0.2], \citet{Dunne2011} [0,0.1], \citet{Marchetti2016} [0.02,0.1], \citet{Wang2016} [0.02,0.1]\\
\hline
$350\mathrm{\mu m}$ & HERSCHEL SPIRE 350 &  354.27&  84.50 & \citet{Vaccari2010} [0,0.2], \citet{Negrello2013} [$\lesssim 0.023$], \citet{Marchetti2016} [0.02,0.1]\\
\hline
$500\mathrm{\mu m}$ & HERSCHEL SPIRE 500 &  515.36&  143.92 & \citet{Vaccari2010} [0,0.2], \citet{Marchetti2016} [0.02,0.1]\\
\hline
$850\mathrm{\mu m}$ & JCMT SCUBA2 850 &  853.81&  88.82 & \citet{Dunne2000} [0.002,0.026], \citet{Vlahakis2005} [0.008,0.024], \citet{Negrello2013} [$\lesssim 0.023$] \\
\hline
TIR & $-$ & $-$ &  $-$ & \citet{Vaccari2010} [0,0.2], \citet{Gruppioni2013} [0,0.3], \citet{Marchetti2016} [0.02,0.1] \\
\hline
\end{tabular}
\label{table:filter}
\vspace{1mm}
\end{table*}

\subsection{Observational data}
\label{sec:obs_data}
To make comparisons between the \colibre broadband LFs and observations, we collect observational data from various previous studies. As this paper focuses only on local LFs, we restrict our observational results (used for comparison) to the measurements with the upper limit of their redshift range below $z=0.3$, a relatively generous threshold chosen to include as many observational datasets as possible\footnote{The measurements from \citet{Moore2026} are based on objects at $z \lesssim 0.6$, but are corrected for evolution to $z = 0.1$. We therefore still include \citet{Moore2026} for comparison in this work.}. For references that provide LFs in multiple redshift bins meeting this criterion (e.g. \citealt{Budavari2005}), however, we only adopt the measurements from their lowest-redshift bins and treat them as local constraints in our plots. A caveat of this selection criterion is that possible cosmic evolution of the observed LFs is not accounted for in our default \colibre comparison, since we only use galaxies from \colibre at $z=0$. This effect may not be negligible, as previous work has shown measurable LF evolution even within the range $z\lesssim 0.3$ \citep[e.g.][]{Loveday2012,Gruppioni2013}. However, because our compilation includes both measurements based on samples extending to $z\approx 0.3$ (such as \citealt{Rodighiero2010,Gruppioni2013}) and studies restricted to much lower redshifts (e.g. \citealt{Hill2010,Dunne2000}), we still adopt the \cosk LFs at $z=0$ as our default reference for comparison, while presenting a test of cosmic evolution on the LFs in Appendix~\ref{sec:LF_z-dependence}. We find that the FIR band shows the largest cosmic evolution, with the luminosities of galaxies at a number density of $10^{-4}\,\rm Mpc^{-3}\,dex^{-1}$ brightening by $\approx 0.4 \,\rm dex$ from $z=0$ to $z=0.3$ (see Appendix~\ref{sec:LF_z-dependence} for more details).

We summarize the observational data used for comparison, together with their redshift ranges, in \autoref{table:filter}. We note that different observational studies adopt different units to express the luminosities of galaxies, for example, absolute magnitudes (e.g. \citealt{Driver2012,Loveday2012}) and monochromatic spectral luminosities multiplied by frequency, $\nu L_{\nu}$ (e.g. \citealt{Rodighiero2010,Negrello2013}). The former are typically used at shorter wavelengths (e.g. FUV to $K$ band), while the latter are commonly used at longer wavelengths. For simplicity, in this work we express all galaxy luminosities as $\nu L_{\nu}$, in units of $\mathrm{L_{\odot}}$, where $\nu$ is the frequency corresponding to the pivot wavelength of the adopted filter and $\mathrm{L_{\odot}}$ is the bolometric luminosity of the Sun ($\mathrm{L_{\odot}} = 3.828 \times 10^{26}\,\mathrm{W}$, following \citealt{Mamajek2015}). Before making comparisons, we transform all galaxy luminosities and the corresponding galaxy number densities in the luminosity bins into the format of $\log\,\nu L_{\nu}$ and $\log\,\Phi(\log\,\nu L_{\nu})$ (i.e., $\Phi$ is defined as the galaxy number density per dex in $\nu L_{\nu}$). For visualization purposes, we thin out the plotted observational measurements when they are too densely spaced in luminosity, retaining at most one point every 0.25 dex on average (i.e. no more than four points per dex in luminosity), in order to avoid overcrowding and improve the readability of the figures.

Differences in the assumed Hubble constant result in systematic differences in the inferred distances of observed galaxies, affecting both luminosities and number densities. To ensure a consistent comparison between our simulations and observational data, we rescale all luminosities in observational data to the same Hubble constant used in {\sc colibre}, using $\nu L_{\nu}\times (h_{\rm sim}/h_{\rm obs})^{-2}$, where $h_{\rm obs}$ is the Hubble constant stated in the references and $h_{\rm sim}$ is the Hubble constant used in \colibre ($h_{\rm sim}=0.681$). In addition, all observed number densities are also scaled to the Hubble constant used in {\sc colibre}, using $\Phi \times (h_{\rm sim}/h_{\rm obs})^{3}$. This standard convention effectively removes the dependence on the Hubble constant from our comparison of simulations with observations and allows for a fair comparison across datasets adopting different cosmologies. We note that differences in other cosmological parameters, such as $\Omega_{\rm m}$, can also affect the inferred LFs through their impact on distance and volume measurements. However, this effect is typically smaller than that introduced by variations in $h$, especially at low redshift, and it is not straightforward to correct for in an analytic way. We therefore do not apply an explicit correction for differences in $\Omega_{\rm m}$ here.

Collecting observational data from multiple studies can be time-consuming and repetitive. To support reproducibility and facilitate future comparisons, we provide all observational datasets we have collected for the comparisons in this series of papers in machine-readable format at \url{https://icc.dur.ac.uk/data/} and on the journal website. We hope that these data will serve as a useful resource for the community and will encourage similar practices in future work.

\begin{figure*}
\centering
\includegraphics[width=0.9\textwidth]{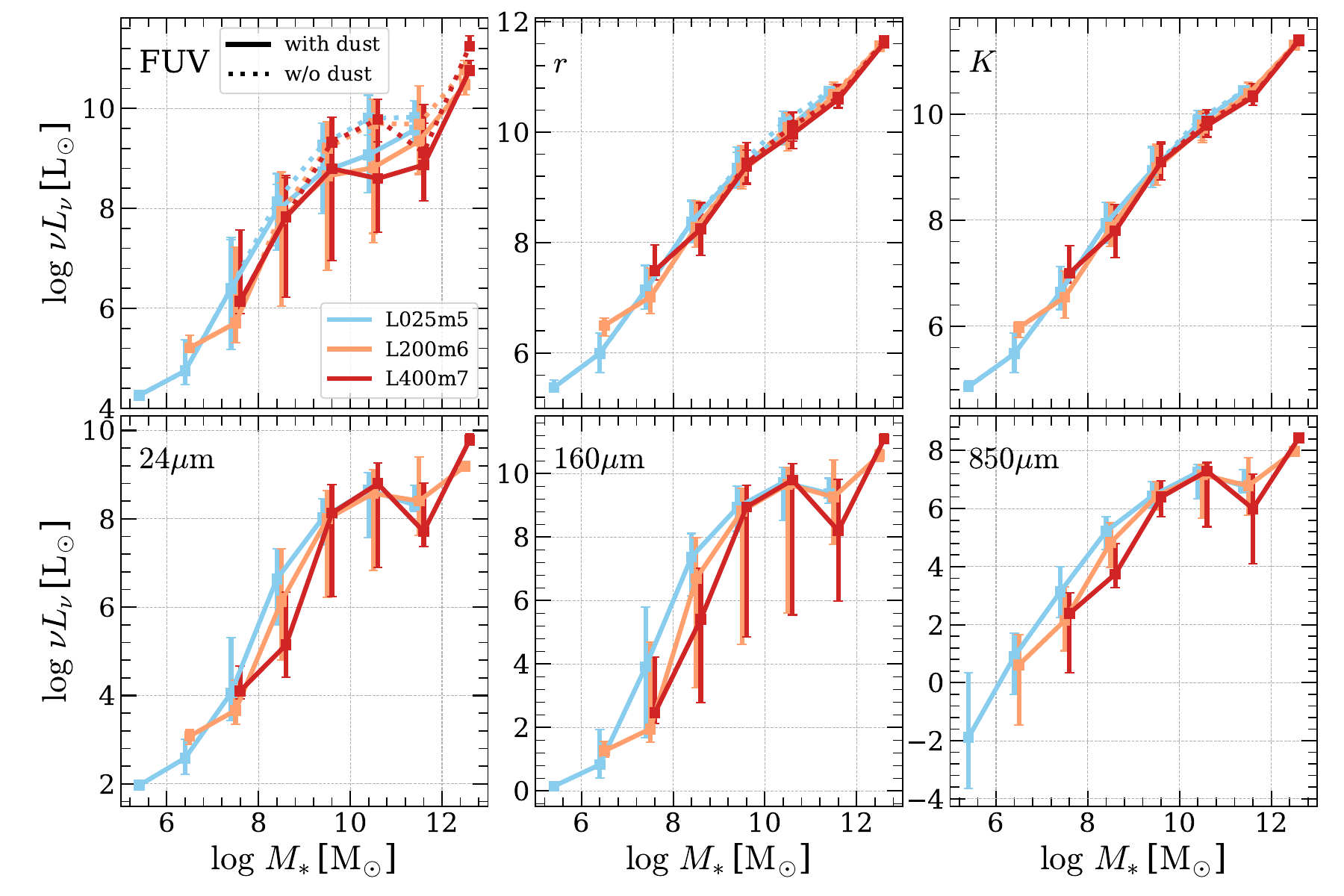}\vspace{-0.2cm}
\caption{The correlation between galaxy stellar mass and monochromatic luminosities (for six example bands, from FUV to 850 \micron) for the selected samples. In each panel, the results from different \colibre simulations are indicated by different colours. The data points represent the median value in each stellar mass bin and the error bars represent the range from the 16th to the 84th percentiles (the $\pm 1\sigma$ range). For the FUV, $r$ and $K$ bands, we show monochromatic luminosities both with dust (solid lines) and without dust (dotted lines), while for the other bands (where dust emission is not negligible) we show only the results with dust (accounting for both dust attenuation and dust emission).}
\label{fig:mstar2lum}
\end{figure*}

\section{Results}
\label{sec:result}
\subsection{Correlations between stellar mass and luminosity}
\label{sec:mstar2lum}

In \autoref{fig:mstar2lum}, we show the correlations between galaxy stellar mass and luminosities (in the form of $\nu L_{\nu}$) for different bands (from the FUV to 850 \micron).

For the $r$ and $K$ bands, where both dust attenuation and dust emission are generally negligible and the luminosities are insensitive to young stellar populations, the three simulations with different resolutions show excellent agreement over a wide stellar mass range, extending down to very low masses. This indicates that the stellar emission in these bands is independent of resolution in the \colibre model.

In the FUV band, the agreement remains generally good, except at the highest stellar mass end. In this regime, the dust-free results (dotted) show better consistency than the dust-attenuated ones (solid), suggesting that the discrepancies are primarily driven by differences in the dust content and spatial distribution in massive galaxies. This is consistent with the resolution dependence seen in the stellar mass - dust mass relation (\autoref{fig:Mstar2Pros}) and may also be related to the fact that the bright end is predicted to evolve more strongly with redshift (see \autoref{fig:zdependence}).

At longer wavelengths (24, 160 and 850 \micron), where dust emission dominates, good agreement is only found at intermediate stellar masses and the high mass end for m5 and m6. At both the low- and high-mass ends, the discrepancies between different simulations can, at least in part, be attributed to the resolution dependence of the dust mass and its clumping properties.

\subsection{Luminosity functions}
In this section, we present the LFs of \colibre galaxies at $z=0$ modelled using the calibration-free {\sc colibre-skirt} pipeline. Specifically, we present the LFs from FUV to the $K$ band ($0.15-2.2\,\micron$) in \autoref{sec:fuv2k}, those in the mid-infrared (MIR; $3.4-24\,\micron$) bands in \autoref{sec:mir}, those in the far-infrared (FIR) and submillimetre bands ($70-850\,\micron$) in \autoref{sec:fir2submm} and those for the total infrared (TIR) in \autoref{sec:tir}.

In this work, LFs are computed in luminosity bins of width 0.25 dex, while the results are reported as galaxy number densities per dex. In each bin, the galaxy abundance is thus calculated as:
\begin{equation}
\label{eq:lf}
\Phi_{\log\,\nu L_{\nu}} = \frac{\sum_i 1/\eta_i}{\delta_{\rm L}\times L_{\rm cube}^3},
\end{equation}
where $\eta_i$ is the sampling ratio (see \autoref{sec:sample}) of the $i$-th galaxy in the bin, $\delta_{\rm L}$ is the bin width ($\delta_{\rm L}=0.25\,\rm dex$) and $L_{\rm cube}$ is the length of one side of the simulation cube. The summation is taken over all galaxies within the given $\log\,\nu L_{\nu}$ bin. We verified that the choice of luminosity bin width only has a minor influence on the LFs.

As mentioned earlier, \colibre provides simulation runs at different resolutions and with different cube sizes. Simulations at high resolution enable us to study the LFs down to the very faint end, while those in large volumes allow us to explore the rare galaxies at the bright end. Thus, in this work, we employ three simulations (L025m5, L200m6, L400m7) to study the LFs at both the bright and faint ends. In addition to the LFs from the three individual simulations, we also combine the LFs from these different \colibre simulations to construct \cosk LFs across a very wide luminosity range in each band. The combination procedure is described below.
\begin{enumerate}
\item Among the three \colibre simulations, we adopt the LFs from L200m6 as the fiducial results for all bands.

\item Starting from a reference point of $\log\,\nu L_{\nu}$ and moving toward the fainter end, we compare the galaxy abundances from L200m6 and the higher-resolution simulation, L025m5. If the difference between the two in logarithmic space exceeds a threshold (0.25 dex) in two consecutive $\log\,\nu L_{\nu}$ bins (which implies that the sample from L200m6 begins to be incomplete), we switch to using the LFs from L025m5 for all subsequent (fainter) bins. We note that here we take the ``knee'' of the LFs as the reference point for starting. While this choice is somewhat arbitrary, the knee of the LFs roughly corresponds to a galaxy abundance of $\approx 10^{-2}\,\mathrm{Mpc^{-3}\,dex^{-1}}$ across all the bands studied in this work, which corresponds to about 160 galaxies per dex in luminosity even in the simulation with the smallest cube size (L025m5). Additionally, the $\log\,\nu L_{\nu}$ of the knee roughly corresponds to a stellar mass of $10^9-10^{10}\,\mathrm{M_{\odot}}$ (see \autoref{fig:mstar2lum}), which contains at least 50 stellar particles even in the simulation with the lowest resolution (L400m7). We therefore do not expect the knee of the LFs to be dominated by poor sampling of either the galaxy population or the stellar component within individual galaxies.

\item Similarly, starting from the knee of the LFs and moving toward the brighter end, we compare the LFs from L200m6 and the simulation with the larger simulation cube, L400m7. Again, if the galaxy abundance difference (in logarithmic space) exceeds 0.25 dex in two consecutive $\log\,\nu L_{\nu}$ bins, we switch to using the LFs from L400m7 for all subsequent (brighter) bins.
\end{enumerate}

We estimate the statistical uncertainties in the \colibre LFs using a two-stage bootstrap procedure. For each simulation, we generate 1000 bootstrap realizations by independently resampling, with replacement, the full parent galaxy population and the subsample that has already been post-processed with {\sc skirt}. Specifically,
\begin{enumerate}
\item First, the full parent galaxy population is resampled with replacement. Since the resampling is performed over the entire parent population, the number of parent galaxies in each stellar mass-SFR bin naturally fluctuates between realizations.
\item Second, the selected galaxies with existing \skirt post-processing results are independently resampled with replacement within each stellar mass-SFR bin, while keeping the number of selected objects in each bin fixed. The luminosities of the resampled galaxies are taken from their original \skirt calculations, and no additional radiative-transfer calculations are performed.
\item For each realization $k$, the sampling fraction of selected galaxies in the $i$-th stellar mass-SFR bin is then recalculated as:
\begin{equation}
\eta_{i}^{k} = \frac{N_{{\rm sel},i}}{N_{{\rm parent},i}^{k}},
\end{equation}
where $N_{\mathrm{parent},i}^{k}$ is the number of galaxies in this stellar mass-SFR bin after the global bootstrap resampling of the parent population, while $N_{\mathrm{sel},i}$ is the number of selected galaxies in each bin, which is fixed between realizations. The LFs are then recalculated with \autoref{eq:lf}.
\item In each band, the uncertainty of an LF in a given luminosity bin is taken to be the standard deviation of the corresponding LF values across the 1000 realizations.
\end{enumerate}
This procedure captures fluctuations both in the abundance of galaxies in the parent population and in the detailed luminosity properties of the finite subsample processed with {\sc skirt}. For the combined LFs, we apply the simulation-combination procedure described above independently to each bootstrap realization and adopt the standard deviation of the resulting 1000 combined LFs as the uncertainty in each luminosity bin. These uncertainties are generally much smaller than the line widths and are therefore not visible in the figures. We consequently do not show error bars, but provide the uncertainties together with the LF values in machine-readable format.

\subsubsection{From FUV to $K$ band}
\label{sec:fuv2k}

\begin{figure*}
\centering
\includegraphics[width=0.83\textwidth]{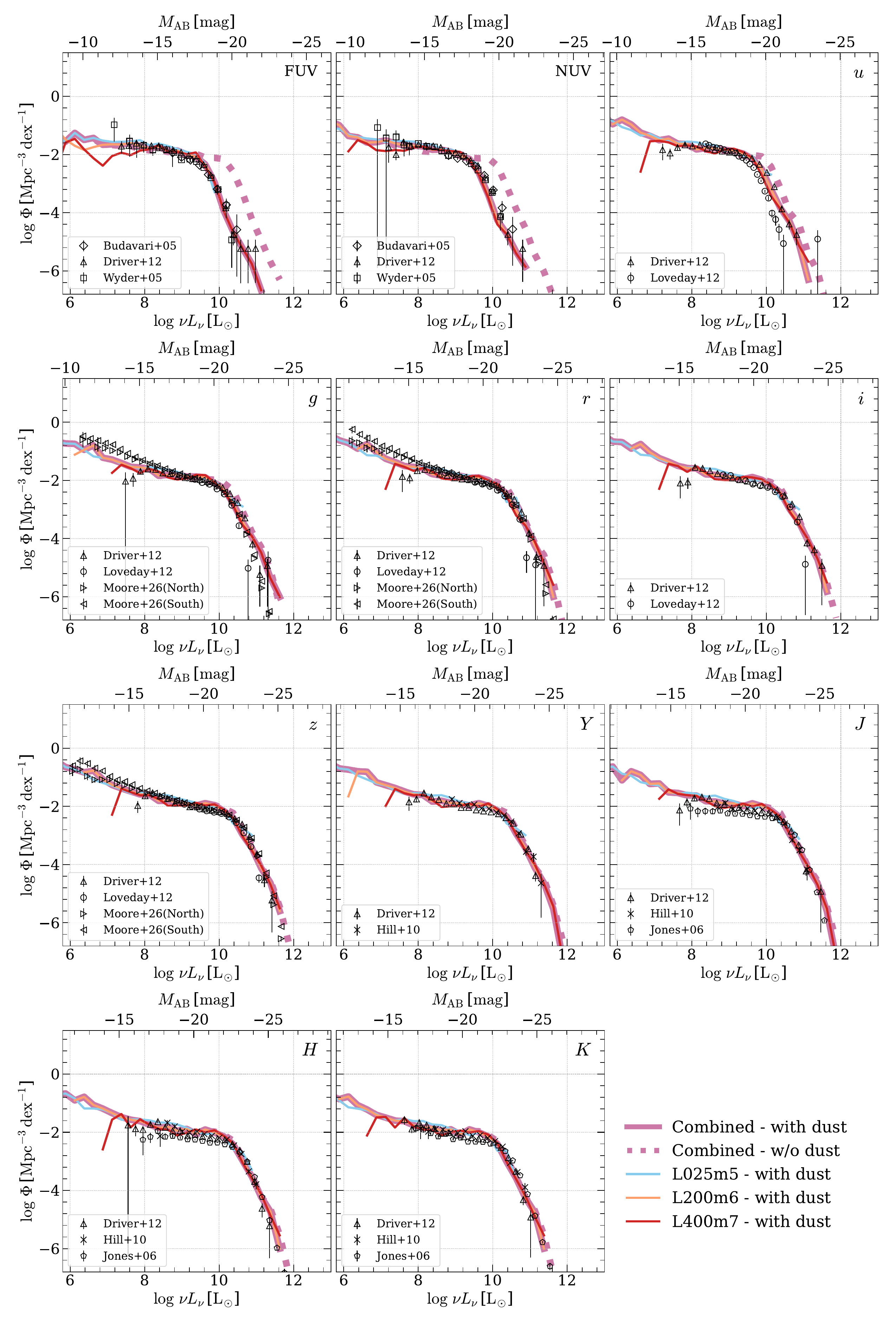}
\caption{\cosk LFs from FUV to $K$ band, compared with observations (see \autoref{table:filter} for the list). In each panel, the LFs from different \colibre simulations (with dust) are shown in different colours. The combined LFs are shown as thick purple curves, with the solid lines indicating results including dust and the dotted lines indicating results without dust. The bottom X-axis of each panel shows the luminosity ($\log\nu L_{\nu}$), while a secondary axis on top indicates the corresponding AB absolute magnitude. A remarkably good agreement of LFs between \cosk and observations is seen.}

\label{fig:FUV2K}
\end{figure*}
In \autoref{fig:FUV2K}, we present the \cosk LFs from three simulations with different resolutions and volume (see \autoref{table:simulations}, with different colours indicating each simulation), together with the combined LFs from FUV to $K$ band (purple thick curves: solid lines show the LFs including dust, dashed lines show those without dust) and compare them to the observational results listed in \autoref{table:filter}. 

\begin{figure*}
\centering
\includegraphics[width=0.7\textwidth]{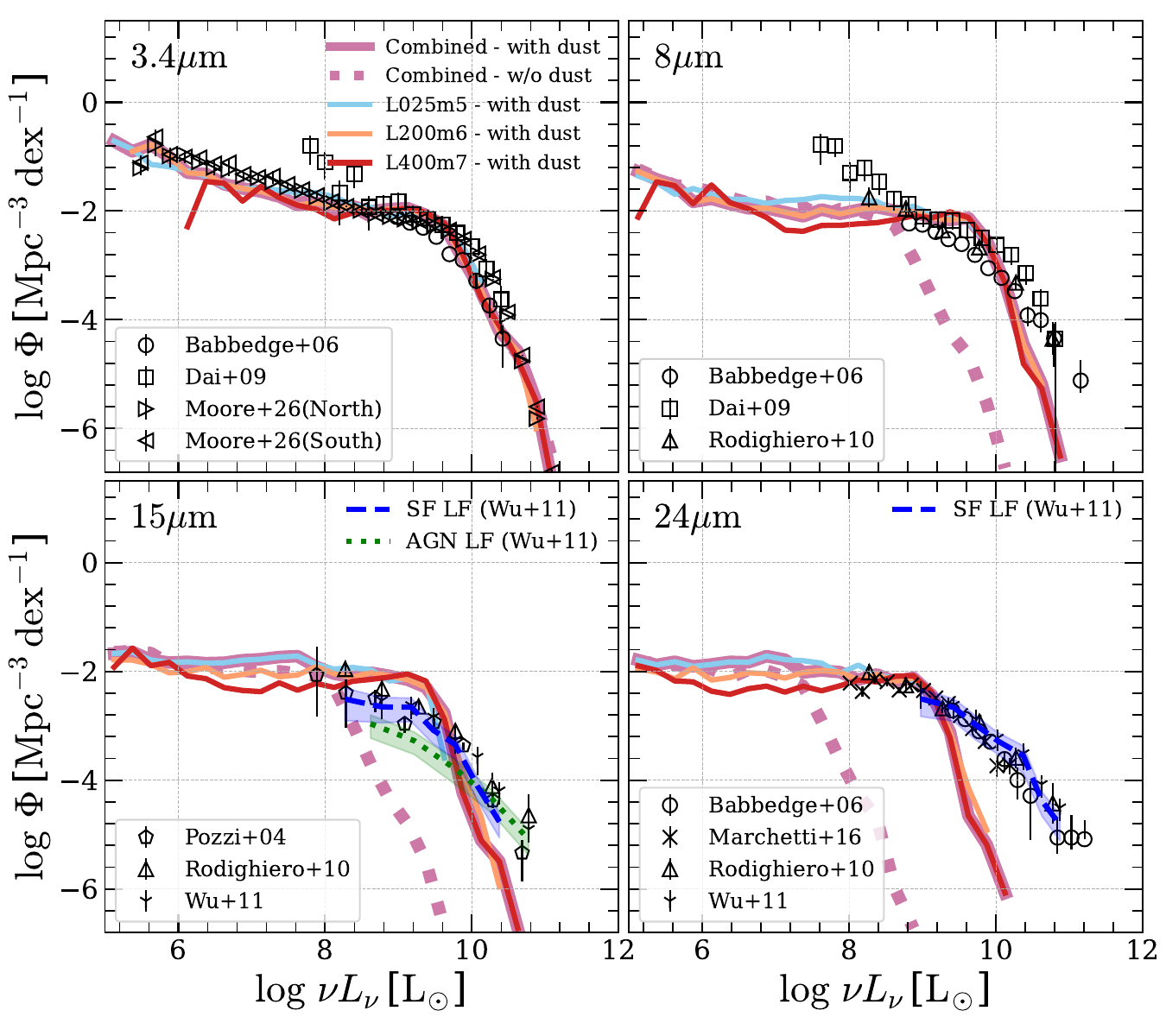}\vspace{-0.2cm}
\caption{Luminosity functions of \colibre in the mid-infrared (MIR) bands (from 3.4 \micron\ to 24 \micron), compared with observations (see \autoref{table:filter}). In the 3.4 \micron\ panel, we also show the observed LF at 3.6 \micron\ from \citet{Babbedge2006} and \citet{Dai2009}. In the bottom panels, the green and blue shaded regions show the observed LFs of the AGN and star-forming components, respectively, from \citet{Wu2011}. The line styles are the same as in \autoref{fig:FUV2K}. \cosk shows good agreement with observations down to very faint end at $3.4\,\mu \rm m$, while it underpredicts the luminosities of galaxies at the bright end for longer wavelengths.}
\label{fig:MIR}
\end{figure*}

\begin{figure*}
\centering
\includegraphics[width=0.85\textwidth]{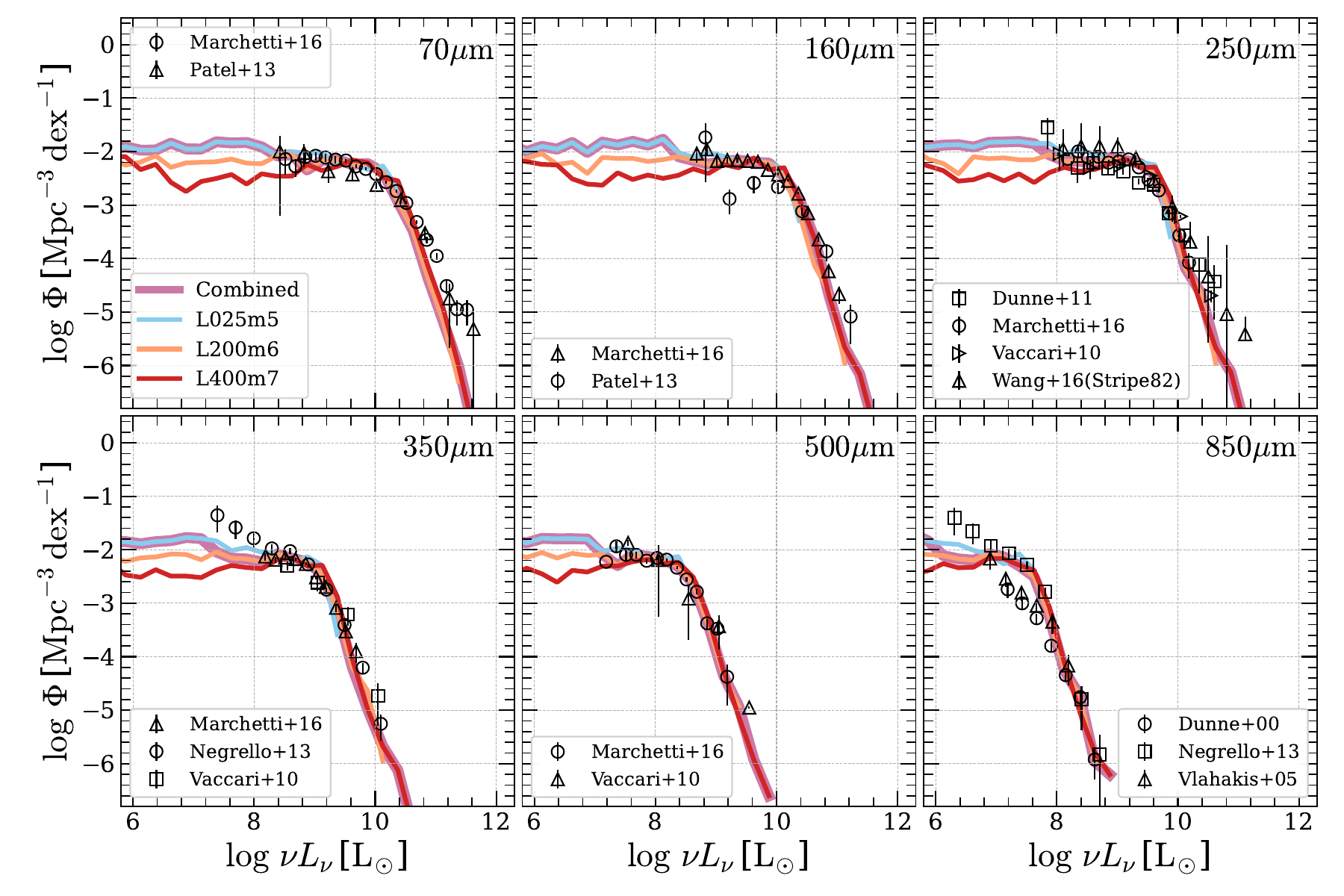}\vspace{-0.2cm}
\caption{Luminosity functions of \colibre in the far-infrared (FIR) and submillimetre bands (from 70 \micron\ to 850 \micron), compared with observations (see \autoref{table:filter}). The line styles are the same as in \autoref{fig:FUV2K}. An overall good agreement of the LFs between \cosk and observations is seen.}
\label{fig:FIR}
\end{figure*}

As can be seen, the LFs from different \colibre simulations of varying resolution exhibit very good agreement over large luminosity ranges. However, below certain luminosities, the LFs of the lower-resolution \colibre simulations begin to deviate from those of the higher-resolution simulations because of faint-end incompleteness caused by the finite numerical resolution. Good convergence in the LFs is also seen (mainly at the bright end) in other bands, such as the MIR bands (\autoref{fig:MIR}), the FIR and submillimetre bands (\autoref{fig:FIR}) and the total IR (\autoref{fig:TIR}). We note that this is not in conflict with the disagreement of the dust mass-stellar mass relation for \colibre simulations at different resolutions (seen in \autoref{fig:Mstar2Pros}). Although the three \colibre simulations show poorer convergence (mainly at the low-stellar-mass end) in that relation, the LFs, especially at the bright end, are dominated by more massive galaxies, for which the agreement between different resolution simulations is better. The poorer convergence of the dust mass-stellar mass relation at low stellar masses instead mainly affects the faint end of the IR LFs, where discrepancies among the three \colibre simulations do indeed become apparent.

When compared to the observational results, we find remarkably good agreement between the combined \cosk (with dust) LFs and the observations across all bands from the FUV to the $K$ band. The only noticeable discrepancy appears in the $ugriz$ bands, especially the $u$ band, where the \cosk LFs agree very well with the results of \citet{Driver2012} (and \citealt{Moore2026} in $grz$ bands\footnote{For the LFs from \citet{Moore2026}, we adopt their Stepwise Maximum Likelihood (SWML) estimates for all available bands.}), while those from \citet{Loveday2012} are generally lower. 
This difference arises from two factors. First, \citet{Loveday2012} $K$-corrected their objects to $z=0.1$ (which is also done in \citealt{Moore2026}), whereas \citet{Driver2012} corrected theirs to $z=0$. Second, \citet{Driver2012} adopted Kron magnitudes, while \citet{Loveday2012} used Petrosian magnitudes. Although the two magnitude definitions agree for late-type galaxies, Petrosian magnitudes progressively miss the outer flux of galaxies that are more centrally concentrated (typically massive and bright systems; e.g. \citealt{Bernardi2013}). According to the comparison presented in \citet{Driver2012}, the latter factor appears to be primarily responsible for the discrepancy, since the first issue was already corrected for in their work.

Differences in the reference redshifts may also affect comparisons in other bands. For example, in the $J$, $H$ and $K$ bands, \citet{Jones2006} used galaxies extending to $z\approx0.2$ to derive the LFs, whereas \citet{Hill2010} and \citet{Driver2012} used only galaxies at $z<0.1$. However, because the median redshift of the \citet{Jones2006} sample is only $0.054$, the differences among these three observational estimates are relatively small. At longer wavelengths, the observational datasets may span different redshift ranges and the possible impact of cosmic evolution should therefore be kept in mind. We discuss this issue in more detail in Appendix~\ref{sec:LF_z-dependence}.

We note here that, except for the unit convention and $h$-correction mentioned in \autoref{sec:obs_data}, we did not apply any corrections to the observational data used for comparison and we interpret the differences arising from their different methods/choices as part of the systematic uncertainties in the observations. For these bands (FUV to $K$), we also show the combined dust-free LFs (thick purple dotted lines) to illustrate the impact of dust. Differences between LFs with and without dust are only apparent at the bright end in the FUV to $u$ bands. This is because dust attenuation is stronger at shorter wavelengths, while dust emission is not yet significant in the FUV-$K$ wavelength range, consistent with theoretical expectations.

A noteworthy result is that the \colibre\ LFs agree with the observations down to the very faint end (as faint as $\log\,\nu L_{\nu}/\mathrm{L_{\odot}}\approx 6$) in the $g$, $r$ and $z$ bands (where measurements from \citealt{Moore2026} based on DESI are available) with only very small differences in the $g$ and $r$ bands. This suggests that \colibre captures the baryonic processes regulating low-mass galaxies reasonably well, including the effects of star formation and feedback, thereby yielding a realistic abundance of faint galaxies and a realistic integrated star formation efficiency in low-mass haloes at $z=0$. We discuss these points further in \autoref{sec:discussion}. This agreement is particularly encouraging because the \colibre subgrid model is calibrated to match observations only over the stellar mass range $10^{9}-10^{11.3}\,\mathrm{M_{\odot}}$. The faint and brightest ends of the LFs therefore probe galaxy populations outside the calibration range. Reproducing the LFs in these regimes is thus not enforced by calibration, but instead represents a genuinely predictive success and an independent test of the model.

\subsubsection{Mid-infrared LFs}
\label{sec:mir}
In \autoref{fig:MIR}, we present the \cosk LFs, including both the combined (with and without dust) results, and those from individual simulations (with dust only), in the MIR bands. 
The \cosk LF with dust (thick solid purple) agrees well with the LF that does not include dust (thick dotted purple) at 3.4 \micron, indicating that the emission at this wavelength is dominated by the stellar component. At longer wavelengths, the bright ends of the LFs without dust are significantly lower than those with dust, reflecting the increasing contribution of dust emission in these bands in intermediate mass and massive objects towards longer wavelengths. 

When comparing the \cosk LFs (with dust) to observational data, we find that the agreement at 3.4~\micron\ remains remarkably good with the data of \citet{Moore2026} down to the very faint end. At 8 \micron, the \cosk LF (with dust) is lower than the observations at both the bright and faint ends. At the faint end, however, we note that the measurements from \citet{Dai2009} may be somewhat high, as their LF is also above that of \citet{Moore2026} at the faint end in the 3.4 \micron\ panel. We therefore refrain from drawing strong conclusions from the faint-end discrepancy between \cosk and \citet{Dai2009}. At longer MIR wavelengths, \cosk predicts systematically fainter galaxy luminosities than observed at the bright end, and thus lower bright-end number densities than the observations. The discrepancy becomes more significant toward longer wavelengths, increasing from $\approx 0.2-0.4$ dex at 8 \micron\ to $\gtrsim 0.8$ dex at 24 \micron\ at a number density of $10^{-4}\,\mathrm{Mpc^{-3}\,dex^{-1}}$. This discrepancy is also seen in \citet{Baes2020}, who studied the IR LFs and dust mass functions in the EAGLE simulations (see their Fig. 1), and can be attributed to several factors, including the absence of an AGN luminosity contribution and the cosmic evolution of the LFs. We will come back to this problem in \autoref{sec:mismatch_mir}.

\subsubsection{Far-infrared and submillimetre LFs}
\label{sec:fir2submm}
The far-infrared (FIR) to submillimetre emission is dominated by thermal radiation from cold dust ($T \approx 20-30\,\rm K$; e.g. \citealt{Draine2003,Galliano2018}) in the ISM. This emission primarily reflects the total dust mass and the radiation field heating the dust, making it a tracer of the dust content and SFRs of galaxies. In \autoref{fig:FIR}, we show a comparison of simulated and observed LFs from 70 to 850 \micron. Given that dust plays a dominant role in the emission in the FIR and submillimetre bands, we present only the LFs with dust (i.e. including dust emission and, although weak, dust attenuation) here, as well as later when we study the total infrared LF.

As can be seen, in the FIR and submillimetre bands, \cosk shows very good agreement with observations\footnote{We note that the observed LF at 850 \micron\ from Table~5 of \citet{Negrello2013} does not match that presented in their Fig.~10, specifically for the bin at $\log L_{353}/\mathrm{W\,Hz^{-1}}=22.12$. We have therefore omitted this bin from our \autoref{fig:FIR}.}, with only a small difference at the brightest end of LF at 70 \micron\ (where \cosk slightly underestimates the abundance of bright galaxies), indicating that the cold dust emission in \colibre closely matches the observations. This further demonstrates that the simulations reliably reproduce the total dust mass and the overall energy balance. This is consistent with the results of \citet{Trayford2026}, who report overall good agreement between the dust mass function in \colibre and observations.

\subsubsection{Total-infrared LF}
\label{sec:tir}
The total infrared (TIR) LF provides a statistical description of the dust-reprocessed energy output of galaxies and serves as a direct tracer of the dust-obscured component of cosmic star formation. Therefore, in addition to the LFs in specific bands, we also present the TIR LF of \colibre and compare it with observations. The sources for the comparison are listed in \autoref{table:filter}.

We follow the approach commonly used in observational studies (e.g. \citealt{Vaccari2010,Gruppioni2013,Marchetti2016}) and calculate the total IR luminosity of galaxies by directly integrating their SEDs (from {\sc skirt}) over the wavelength range 8 to $1000\,\mathrm{\mu m}$. The luminosity in this case is not expressed as $\nu L_{\nu}$ as before, but as an integrated quantity, denoted by $L_{\rm TIR}$. The TIR luminosities are expressed in units of $\mathrm{L_{\odot}}$, adopting $\mathrm{L_{\odot}} = 3.828\times 10^{26}\,\mathrm{W}$ for the bolometric luminosity of the Sun, following \citet{Mamajek2015}. 

\begin{figure}
\centering
\includegraphics[width=0.45\textwidth]{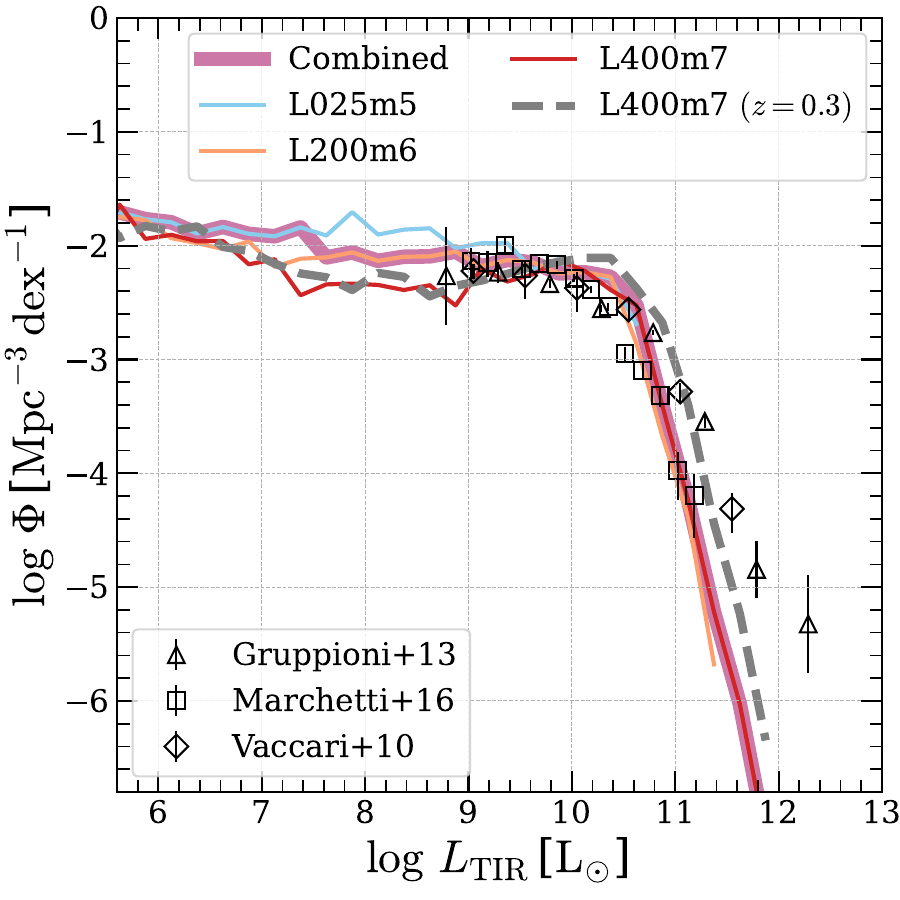}\vspace{0cm}
\caption{The comparison of the total infrared (TIR; $8-1000\,\rm \mu m$) LF between \cosk and observations at $z=0$ (see \autoref{table:filter}). The line styles are the same as \autoref{fig:FUV2K}. The TIR LF at $z=0.3$ in L400m7 is shown by the grey dashed curve. \cosk underpredicts the TIR luminosity of galaxies at the bright end ($\log\, L_{\rm TIR}/\mathrm{L_{\odot}}\gtrsim 11$).}
\label{fig:TIR}
\end{figure}

In \autoref{fig:TIR}, we show a comparison of the TIR LF in \colibre with observations. The infrared SEDs used to derive total IR luminosities from observations are generally not directly sampled over the full wavelength range of interest ($8-1000\,\micron$). Instead, SED models are fitted to photometry in a limited number of bands and the total IR luminosity is then obtained by integrating the best-fitting SEDs. \citet{Rodighiero2010} derived total IR luminosities from SEDs extending only to $24\,\micron$, while \citet{Patel2013} used a sample of galaxies in which only a small fraction has data at $160\,\micron$, and most have data only to $70\,\micron$. We therefore do not include these two studies in our main comparison here, although their results are broadly consistent with the other observational results (e.g. \citealt{Vaccari2010,Gruppioni2013,Marchetti2016}).

For {\sc colibre}, as before, we show results (including dust) from individual simulations at different resolutions, as well as the combined LF. As can be seen, the \cosk\ TIR LF agrees well with the observations from \citet{Marchetti2016} up to $\log\, L_{\rm TIR}/\mathrm{L_{\odot}}\approx 11.2$. Compared to the data of \citet{Vaccari2010} and \citet{Gruppioni2013}, however, the \cosk\ TIR LF shows good agreement only up to $\log\, L_{\rm TIR}/\mathrm{L_{\odot}}\approx 11$, but at the bright end, the \colibre\ galaxies appear not to be bright enough, leading to an underprediction of the TIR LF.

The bright-end deficit in the TIR LF should not be regarded as an independent discrepancy. Instead, it reflects the combined effects of the luminosity deficits already seen in the MIR bands, possibly due to the omission of AGN-heated dust emission and cosmic evolution. Firstly, as shown in \autoref{fig:MIR}, the \cosk LFs are lower than observed in the MIR bands ($8-24\,\mathrm{\mu m}$). Because the TIR luminosity is obtained by integrating the infrared emission over a broad wavelength range, the underestimation of the MIR luminosities naturally propagates into the TIR luminosities and therefore contributes to the bright-end deficit in the TIR LF.

Secondly, although infrared emission powered by AGN is generally strongest at MIR wavelengths and declines towards the FIR, it can still make a non-negligible contribution to the integrated TIR luminosity, particularly in luminous AGN-hosting systems (e.g. \citealt{Mullaney2011}). Consistently with this, \citet{Gruppioni2013} found that AGN-dominated populations become increasingly important towards the bright end of the local TIR LF and dominate the number density at $\log\,L_{\rm TIR}/\mathrm{L_{\odot}}\gtrsim 12$ (see their Fig.~10). The omission of AGN emission in our modelling may therefore lower the predicted TIR luminosities not only through the MIR deficit, but also through the absence of AGN-heated dust emission across the broader infrared SED, particularly for the most luminous systems.

Thirdly, \citet{Vaccari2010} and \citet{Gruppioni2013} calculated their TIR LFs using galaxies up to $z=0.2$ and $z=0.3$, respectively, whereas the \cosk TIR LF is calculated using only $z=0$ galaxies. This may also contribute to the lower TIR LF compared to observations, due to evolution of the LF, which is not accounted for in either the theoretical predictions or observational data plotted in \autoref{fig:TIR}. To test whether cosmic evolution of the TIR LF contributes to the difference, we also present the \cosk\ TIR LF at $z=0.3$ from L400m7, shown as a grey dashed curve in \autoref{fig:TIR}. The $z=0.3$ TIR LF is indeed closer to those from \citet{Vaccari2010} and \citet{Gruppioni2013}, but remains low for $\log\, L_{\rm TIR}/\mathrm{L_{\odot}}\gtrsim 11.5$, indicating that the redshift evolution of the TIR LF predicted by the model cannot explain all of the discrepancy (see Appendix~\ref{sec:LF_z-dependence} for further discussion of cosmic evolution in the LFs).

Finally, differences among the observational estimates also contribute to the apparent discrepancy. Among the three observational datasets shown in \autoref{fig:TIR}, \citet{Marchetti2016} provide the largest survey area and the broadest IR wavelength coverage, whereas the estimates by \citet{Gruppioni2013} and \citet{Vaccari2010} are based on smaller areas and/or more limited IR photometry. The smaller survey areas may make these datasets more susceptible to cosmic/sample variance, especially at the bright end where the number density of galaxies is low. This, together with differences in wavelength coverage and LF estimation, may partly explain why the bright-end discrepancy in the total IR LF is mainly relative to the estimate of \citet{Gruppioni2013} and \citet{Vaccari2010} rather than to that of \citet{Marchetti2016}, which is well reproduced by {\sc colibre}-{\sc skirt}.

\section{Discussion}
\label{sec:discussion}
\subsection{The mismatch in the MIR}
\label{sec:mismatch_mir}

\begin{figure*}
\centering
\includegraphics[width=0.85\textwidth]{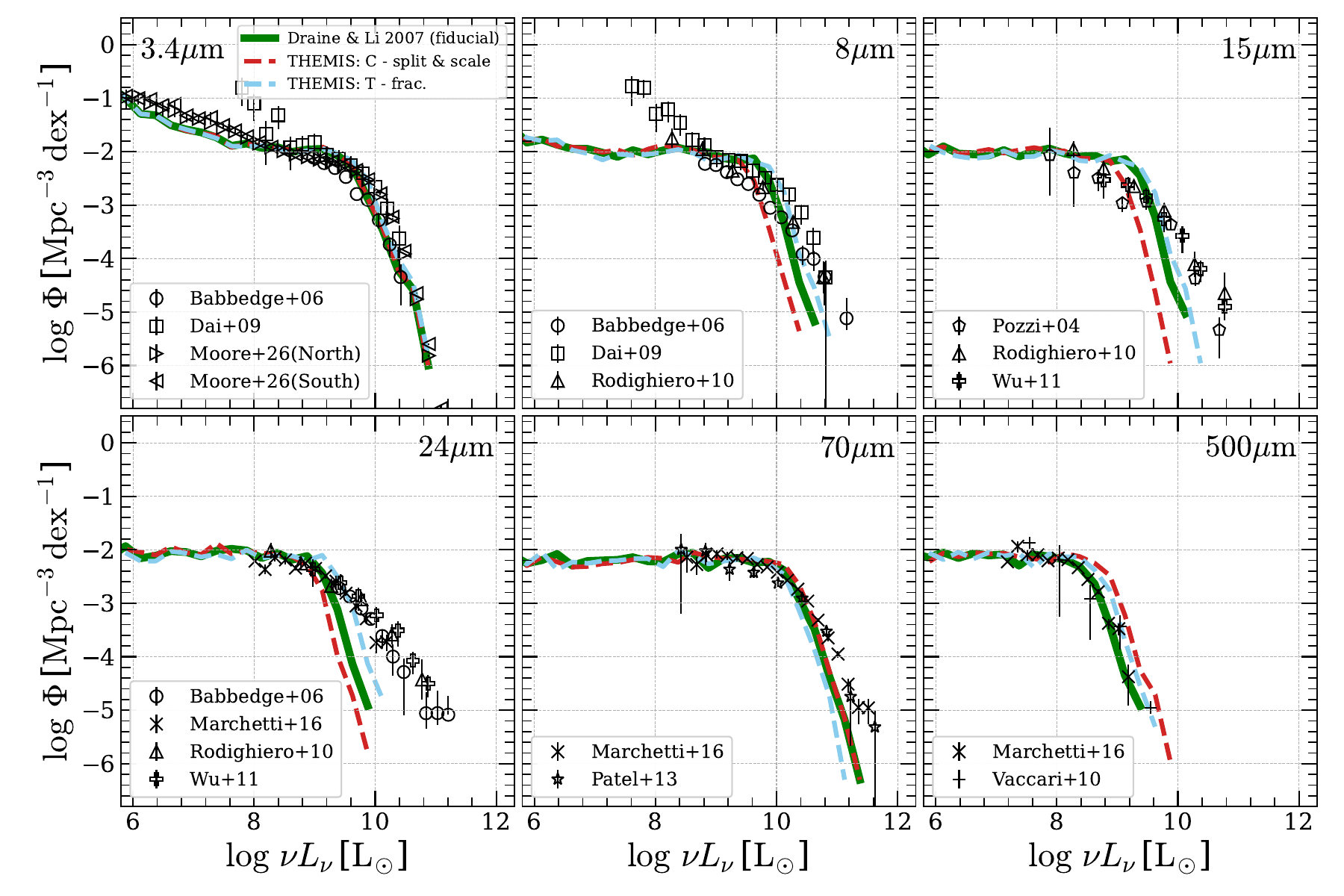}\vspace{-0.2cm}
\caption{Effect of dust models on the LFs in different IR bands. In each panel, the LF obtained with our fiducial dust model from \citep{Draine2007a} is shown as a green solid curve, while the LFs based on two different THEMIS models (\citealt{Jones2017}; see \autoref{sec:mismatch_mir} for details) are shown as red and blue dashed curves. This test is carried out using the \colibre simulation L200m6.}
\label{fig:test_dust_model}
\end{figure*}

\begin{figure*}
\centering
\includegraphics[width=0.85\textwidth]{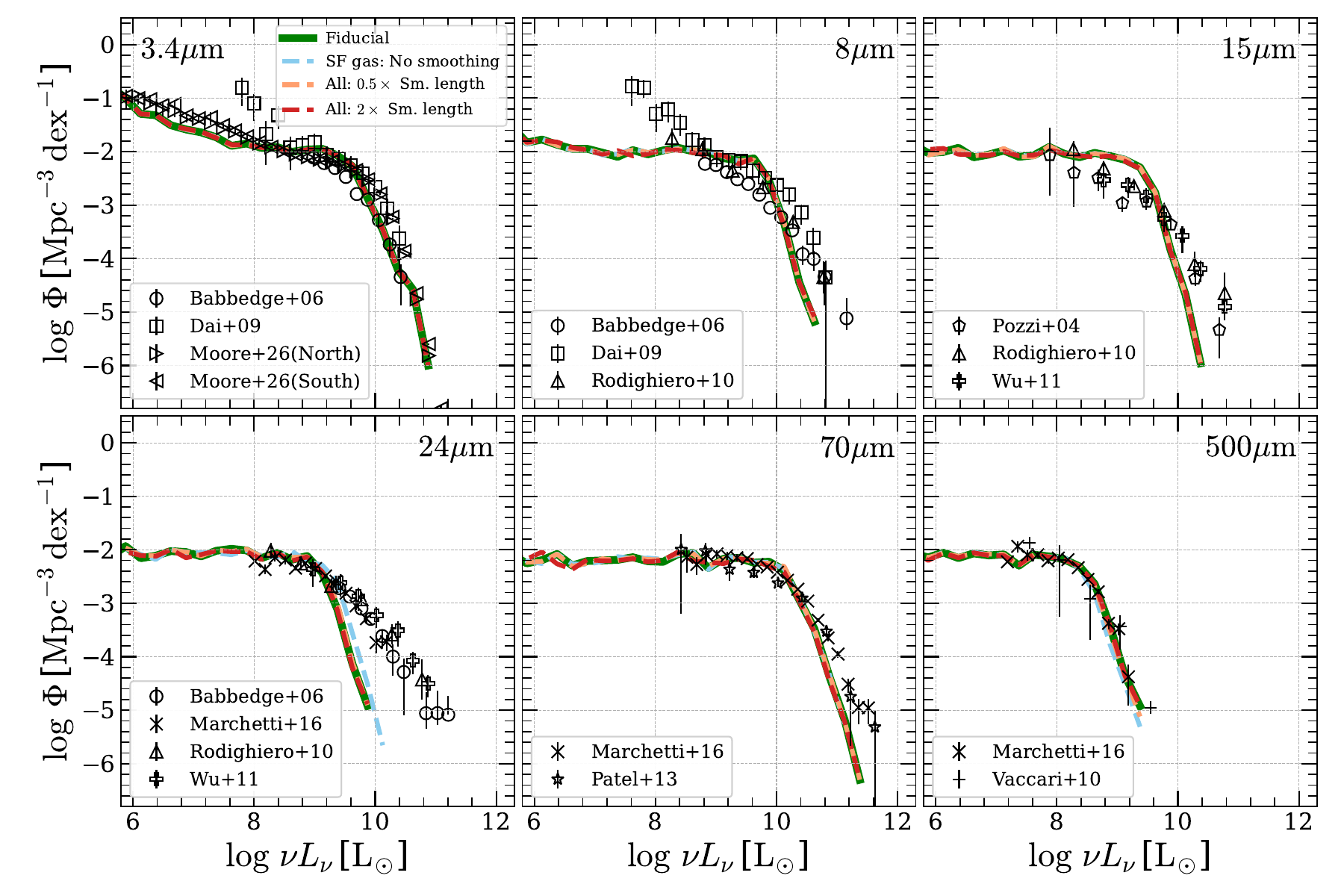}\vspace{-0.2cm}
\caption{Effect of the smoothing lengths of stars and gas on the LFs in different IR bands. In each panel, the fiducial LF is shown as a green solid curve. The LFs obtained by scaling the smoothing lengths of all stellar and gas particles by the specified factors, or by setting those of star-forming gas to 0.001 pc (effectively no smoothing), are shown as dashed curves in different colours. This test is carried out using the \colibre simulation L200m6.}
\label{fig:test_smoothing}
\end{figure*}

\begin{figure*}
\centering
\includegraphics[width=0.85\textwidth]{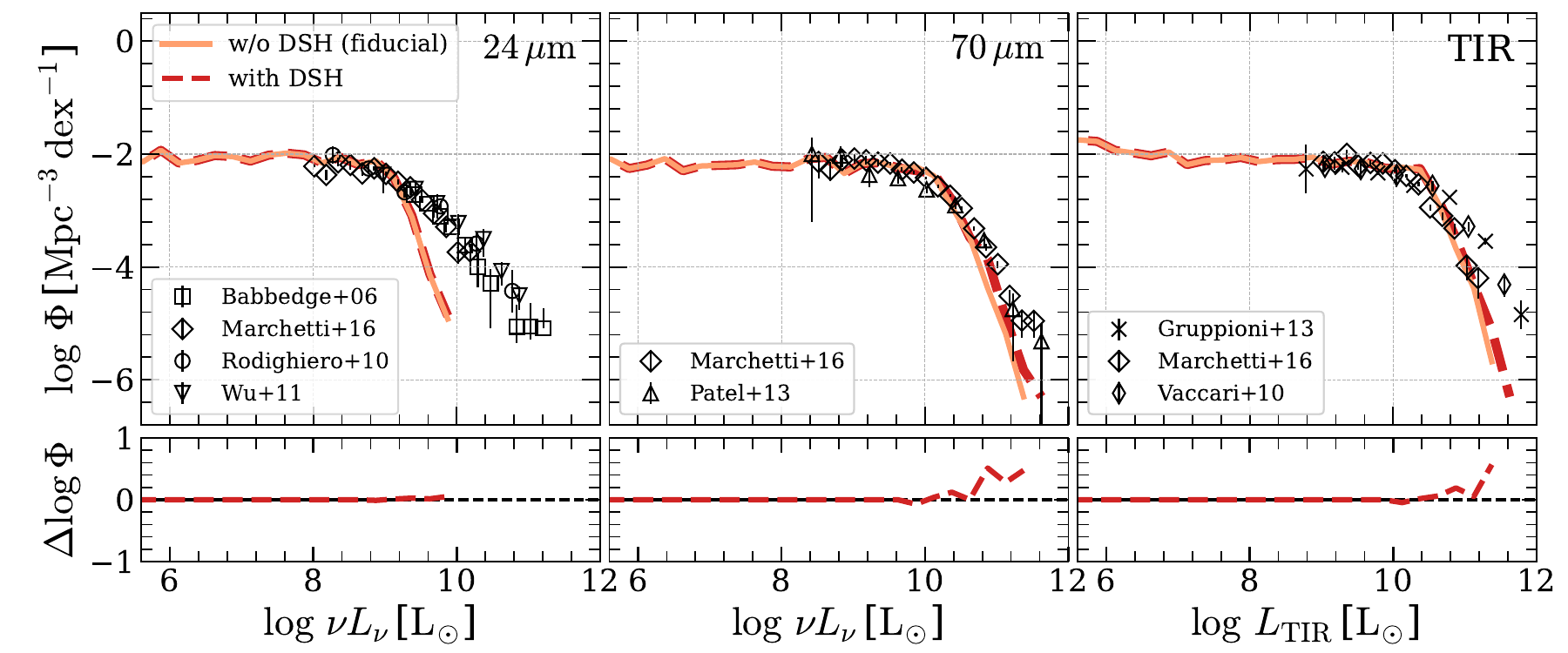}\vspace{-0.2cm}
\caption{Effect of dust self-heating (DSH) on the MIR ($24\,\mu{\rm m}$), FIR ($70\,\mu{\rm m}$), and TIR LFs. The upper subpanel of each panel compares the fiducial LF without DSH (orange solid curve) with the approximate DSH-corrected LF (red dashed curve), constructed by replacing the luminosities of galaxies with fiducial $L_{\rm TIR}\geqslant 10^{10}\,\mathrm{L_\odot}$ with those obtained with DSH enabled. The lower subpanel shows the difference between the two LFs, defined as $\Delta\log\Phi\equiv\log\Phi_{\rm DSH-corrected}-\log\Phi_{\rm fiducial}$, as a function of luminosity.}
\label{fig:test_dsh}
\end{figure*}

As can be seen in \autoref{fig:MIR}, \cosk underestimates the bright end of the LFs compared to observations in the MIR bands at wavelengths $\sim 8-24\,\mathrm{\mu m}$, with the strongest tension occurring at $24\,\mathrm{\mu m}$. Emission over this wavelength range is produced by several different physical components of interstellar dust. At $\approx 5-20\,\mathrm{\mu m}$, the emission is strongly influenced by PAH features. At longer MIR wavelengths, especially around $24\,\mathrm{\mu m}$, it is more closely linked to warm dust heated by young stars and, in some cases, circumstellar dust around evolved stars. In addition, emission from dust heated by AGN, which is not included in our current \skirt calculations, can also be important. Given this complexity, the systematic underestimation across this wavelength range likely reflects a combination of modelling uncertainties and physical limitations in the current framework, rather than a single identifiable cause. In the following, we therefore assess several plausible sources of uncertainty in our modelling. These tests are intended to illustrate the sensitivity of the predicted LFs to different assumptions, rather than to provide a definitive resolution of the discrepancy.

\begin{enumerate}
\item As mentioned in \autoref{sec:skirt}, the abundance of PAH molecules is not modelled self-consistently in {\sc colibre}, but is instead added in post-processing using \skirt with the rescaled \citet{Draine2007a} dust model. The strength of the PAH emission therefore depends on the assumptions adopted in the dust model, particularly the assumed PAH abundance. To assess the impact of the adopted dust model, in \autoref{fig:test_dust_model}, we compare the LFs obtained with two different dust models, namely those by \citet{Draine2007a} and THEMIS \citep{Jones2017}, in the $8-24\,\micron$ bands, as well as in the $3.4\,\micron$, $70\,\micron$ and $500\,\micron$ bands for reference. For THEMIS, we adopt two variations. The first adopts the same ``split \& scale'' approach (see \autoref{sec:skirt} for details) as for the \citet{Draine2007a} model and uses the dust-species fractions from \colibre (referred to as THEMIS: C - split \& scale). The second adopts only the total dust mass from {\sc colibre}, while the relative fractions of the different dust species are taken from the THEMIS model itself (referred to as THEMIS: T - frac). As can be seen, compared to the fiducial model, the THEMIS: C-split \& scale model reduces the luminosities of galaxies that are bright in the $8-24\,\micron$ bands, while increasing the luminosities of galaxies that are bright at $500\,\micron$, consistent with the changes in the cosmic spectral energy distribution (CSED) when adopting the same THEMIS variation shown in \citet{Gebek2026}. By contrast, the THEMIS: T-frac model increases the luminosities of galaxies that are bright in the $8-24\,\micron$ bands and at $500\,\micron$, while slightly reducing the luminosities of galaxies that are bright at $70\,\micron$. 

\item In the fiducial \cosk pipeline, smoothing lengths are used to map discrete star and star-forming gas particles onto a continuous density field. A smaller smoothing length therefore typically results in a more concentrated radiation field around stellar sources and may heat the surrounding dust more effectively, potentially influencing the LFs in the IR bands. In \autoref{fig:test_smoothing}, we present a comparison of LFs obtained with different smoothing length variations in the MIR bands, as well as in the $3.4\,\micron$, $70\,\micron$ and $500\,\micron$ bands for reference. The variations include (1) halving the original smoothing lengths of both evolved stars and star-forming gas, (2) doubling the original smoothing lengths of both evolved stars and star-forming gas and (3) effectively no smoothing for star-forming gas, with the smoothing lengths set to 0.001 pc (the smoothing lengths of the evolved stars remain the original values). Readers are referred to \citet{Gebek2026} for the details of the calculation of smoothing lengths. As can be seen, at $24\,\micron$, the galaxies without smoothing are indeed brighter than those with smoothing by $\approx 0.1$ dex, while at $500\,\micron$, galaxies without smoothing are slightly fainter, consistent with the changes in the CSED when no smoothing lengths are adopted, as shown in \citet{Gebek2026}. For the other bands, however, the variations in smoothing lengths do not make a measurable difference to the LFs.

\item Cosmic evolution of the LFs may also contribute to the discrepancy, as galaxies are observed to be, on average, brighter in the MIR at even slightly higher redshifts (e.g. \citealt{LeFloch2005,Gruppioni2013,Patel2013}). As shown in Appendix~\ref{sec:LF_z-dependence}, this effect makes the galaxies brighter by $0.2-0.4$ dex at the bright end of the MIR LFs.

\item Dust in the obscuring tori around AGN that is directly heated by radiation from the AGN is known observationally to dominate the total MIR emission from some galaxies \citep[e.g.][]{Wu2011}; it is not included in the current {\sc colibre-skirt} pipeline. However, the observational MIR and FIR LFs with which we compare generally include all of the IR emission from a galaxy, including any AGN component. Empirical AGN infrared SEDs show that AGN emission can be particularly strong at MIR wavelengths; therefore the absence of AGN emission in our model may contribute to the discrepancy between the model predictions and the observations. Radiative transfer models of AGN dust torus emission predict that its strength and spectral shape depend on the geometry, optical depth, covering factor and viewing angle of the torus \citep{Mullaney2011,Stalevski2012,Stalevski2016}. The AGN luminosity function predicted by {\sc colibre} will be the topic of a future study, so here we instead estimate the contribution of AGN torus emission to the MIR LFs observationally.

In the bottom panels of \autoref{fig:MIR}, we show observational estimates of the AGN and star-forming LFs at 15~\micron\ and of the star-forming LF at 24 \micron\ at $z \sim 0$ from \citet{Wu2011} (they do not provide the AGN LF at 24 \micron). These LFs are based on a MIR-selected sample of galaxies at $z<0.3$, for which the contributions of the AGN torus and the host galaxy (i.e. dust heated by stars) to the 15 and 24~\micron\ luminosity were separated using MIR spectroscopy. As shown in the figure, at 15 \micron, the star-forming LF from \citet{Wu2011} agrees more closely with the \cosk LFs than estimates that do not remove the AGN contribution. At a fixed number density, for example $\log\,\Phi/(\mathrm{Mpc^{-3}\,dex^{-1}})\approx -5$, the luminosity difference between the \cosk LF and the star-forming LF from \citet{Wu2011} is $\approx 0.4$ dex, compared with $\approx 0.8$ dex when the total LF (i.e. including both AGN and star-forming components) from \citet{Wu2011} is used for the comparison. At 24 \micron, however, the AGN contribution is much smaller than at 15 \micron, resulting in a star-forming LF that lies close to the total 24 \micron\ LF, with a horizontal difference of only $0.1-0.2\,\rm dex$ at $\log\,\Phi/(\mathrm{Mpc^{-3}\,dex^{-1}})\approx -5$. This comparison suggests that including AGN emission could account for a substantial fraction of the discrepancy between the \cosk predictions and observations at 15 \micron, but would have a much smaller effect at 24 \micron.

\item The current fiducial \cosk model does not include dust self-heating (DSH), i.e. the re-absorption of radiation emitted by dust is not taken into account when calculating the dust temperature distribution. Neglecting DSH reduces the computational cost of the radiative-transfer calculation, but may lead to an underestimate of the infrared emission, particularly for galaxies with high dust optical depths. To assess this effect, we select a subsample of the L200m6 selected sample and re-run \cosk with DSH enabled. We include all galaxies with fiducial $L_{\rm TIR}\geqslant 10^{10}\,\rm L_{\odot}$ and randomly select 10 per cent of the galaxies in each $L_{\rm TIR}$ bin below this threshold, resulting in a subsample of 655 galaxies from the original sample of 2994 galaxies. We compare the luminosities obtained with and without DSH from the MIR to the submillimetre, as well as the TIR luminosities. We find that DSH has a negligible effect on low-luminosity galaxies in all of the wavelength ranges considered.

We then construct an approximate DSH-corrected LF by replacing the fiducial luminosities of galaxies with fiducial $L_{\rm TIR}\geqslant 10^{10}\,\rm L_{\odot}$ with those obtained with DSH enabled, while retaining the fiducial luminosities of all galaxies below this threshold. As shown in \autoref{fig:test_dsh}, including DSH has little effect on the MIR ($24\,\mathrm{\mu m}$) luminosities, but increases the FIR ($70\,\mathrm{\mu m}$) and TIR luminosities of the brightest galaxies by approximately $0.1-0.2$ dex. Owing to the steep bright-end slopes of the LFs, these luminosity shifts increase the predicted number densities by up to approximately $0.5$ dex, bringing the \cosk predictions into even closer agreement with the observations.

\item Our current models do not include either emission from sub-grid dust in star-forming regions or nebular continuum emission from ionized gas in these regions. This results from our use of the dust-free templates from the {\sc toddlers v1} library, for the reasons outlined in Section~\ref{sec:skirt}. Dust within star-forming regions would be heated much more strongly than dust in the general ISM, and so could make an important contribution in the MIR. Nebular continuum emission can dominate over intrinsic stellar emission at MIR wavelengths for young stellar populations \citep{Byler2017}, and so might possibly contribute significantly to the total MIR luminosity in some galaxies. We plan to explore both of these effects in future work using the {\sc toddlers v2} library, which allows more flexibility in the choice of sub-grid dust content in star-forming regions as well as in the nebular emission.
\end{enumerate}

Overall, these tests demonstrate that the predicted MIR LFs are quite sensitive to several aspects of the modelling. Although most of the individual variations produce relatively modest changes, their combined effect, if they all act in the same direction, could plausibly shift the bright end towards higher luminosities by $\approx 0.5-0.9$ dex at the brightest end. This estimate includes increases of $0.2-0.4$ dex from adopting the THEMIS: T-frac model (\autoref{fig:test_dust_model}), $\approx 0.1$ dex from removing smoothing for star-forming gas (\autoref{fig:test_smoothing}), and $0.2-0.4$ dex from accounting for the cosmic evolution of the LFs (\autoref{fig:zdependence}). This range is comparable to the model-data discrepancy in the MIR, which, at $\log\,\Phi/(\mathrm{Mpc^{-3}\,dex^{-1}})\approx -5$, is about 0.4 dex at 15 \micron\ and 1.0 dex at 24 \micron\ when the AGN contribution is removed from the observational LFs. Possible contributions from sub-grid dust and nebular continuum emission are also not included in this estimate. The effect of dust self-heating is also not included in this estimate because, as shown in \autoref{fig:test_dsh}, it has a negligible effect on the 24 \micron\ LF. This estimate should, however, be regarded as illustrative rather than quantitative. The effects considered here should not be interpreted as independent or strictly additive and most of them provide only approximate indications of their possible magnitude rather than fully self-consistent predictions.

\subsection{Comparison with previous studies}
\label{sec:comparison}
Many efforts have been devoted to comparing the multi-band LFs of both semi-analytic models and cosmological simulations with observations. For example, \citet{Granato2000} combined the semi-analytic model {\sc galform} \citep{Cole2000} with the radiative transfer code {\sc grasil} \citep{Silva1998} to study local multi-band LFs and compare them with observations. While they found good agreement with the observed local LFs in several bands, including the UV, $B$ and $K$ bands and the IR at $12$--$100\,\rm \mu m$, this agreement should be interpreted with caution. The underlying model was calibrated using local galaxy constraints, including the $B$-band LF, while parameters affecting the infrared emission, such as the burst star formation timescale and some dust-related parameters, were also tuned against local data. Therefore, the agreement does not provide a fully independent test of the model and cannot by itself be taken as strong evidence for its predictive power. \citet{Lagos2019} carried out a similar study using the {\sc shark} semi-analytic model \citep{Lagos2018}, which was calibrated to reproduce the stellar mass functions at $z=0$, 1 and 2, the local black hole-bulge mass relation and the galaxy mass-size relations, rather than the local LFs themselves. They found that the model reproduces the observed $z\approx 0$ LFs from the FUV to the FIR reasonably well overall, although an overabundance in the brightest bins is apparent in many bands, particularly in the IR.

Similar studies have also been carried out using hydrodynamic simulations. For example, \citet{Trayford2015} employed stellar population synthesis modelling to investigate the $u$ to $K$ band LFs of the EAGLE simulations and reported overall good agreement with observations, but did not consider the MIR/FIR bands, where dust emission becomes important. Using the {\sc skirt} radiative transfer code, \citet{Baes2019} studied the CSED of EAGLE and found agreement with observational data to better than 0.15 dex over the full wavelength range at $z=0$, except in the UV, where the model overestimates the UV emissivity by up to a factor of two. Focusing on the infrared, \citet{Baes2020} investigated the {$z\leqslant 0.2$} IR LFs of EAGLE using the same approach and reported a generally lower abundance of galaxies in bands from 8 to 350 \micron\ than observed. For the IllustrisTNG simulations, \citet{Trcka2022}, also using {\sc skirt}, analysed the UV to submillimetre LFs of TNG50 and found an overall excess of galaxies in most bands compared to observations, unless a relatively small aperture of 10~kpc for simulated galaxies was adopted. In particular, they found that both their fiducial aperture ($5\,R_{1/2}$, where $R_{1/2}$ is the three-dimensional stellar half-mass radius of the galaxy) and a 30~kpc aperture result in a luminosity excess of $\sim 0.6-0.8$~dex at a galaxy abundance of $\log\,\Phi/\mathrm{Mpc^{-3}\,dex^{-1}} = -4$, which is not seen in our work. A similar analysis by \citet{Gebek2024} for TNG100 yielded significantly better agreement than that found for TNG50; they attributed the large discrepancy between the TNG50 results of \citet{Trcka2022} and the observations to resolution-dependent differences in the simulation, arising because the IllustrisTNG subgrid model parameters were calibrated at the TNG100 resolution and were not recalibrated for TNG50. However, although \citet{Gebek2024} showed better agreement with observations than \citet{Trcka2022}, their simulated $u$ to $K$ band LFs remained generally higher than observed, particularly at the bright end. This suggests that the self-consistent modelling of the cold ISM and dust attenuation in {\sc colibre} improves the agreement between the predicted and observed LFs. In addition, \citet{Trcka2022} and \citet{Gebek2024} both adopted a \citet{Kroupa2001} IMF for star-forming regions and a \citet{Chabrier2003} IMF for evolved stars, whereas our pipeline consistently adopts a \citet{Chabrier2003} IMF throughout. It is also worth noting that the {\sc skirt} dust post-processing adopted in these studies was not fully calibration-free. In EAGLE-{\sc skirt}, the free parameters were calibrated against local submillimetre colours and dust scaling relations, while in the TNG50/TNG100 analyses the radiative transfer setup was calibrated through comparisons with DustPedia-based \citep{Davies2017} luminosity, colour and SED relations.

In contrast, the \colibre simulations when post-processed with \skirt exhibit remarkably good agreement with the observed $z=0$ LFs over an exceptionally broad wavelength range, spanning from the FUV to $3.4\,\micron$ and from 70 to $850\,\micron$. To our knowledge, this is the first time that such a high level of agreement has been achieved across such an extended wavelength range within a single cosmological simulation framework, using a radiative transfer pipeline that is calibration-free. Here, by ``calibration-free'', we refer specifically to the {\sc skirt} post-processing, which was not tuned to match the observed LFs or other light-based constraints, apart from the choice of dust model: among the dust models tested, the \citet{Draine2007a} model is adopted because it gives better agreement with the observed CSED, while this has only a minor effect on the LFs, as shown in \autoref{fig:test_dust_model}. No other {\sc skirt} parameters were calibrated. This differs from several previous studies (e.g. \citealt{Baes2020,Trcka2022,Gebek2024}). The underlying {\sc colibre} dust model does include a free clumping factor used to account for unresolved dense gas when calculating dust and H$_2$ contents, but this parameter was not calibrated against any particular dust dataset and the low-redshift dust content is largely insensitive to its precise value once it exceeds a minimum threshold (see \citealt{Trayford2026} for details). 

This result is particularly noteworthy because {\sc colibre} itself was calibrated only to reproduce the observed stellar mass function, the stellar mass-size relation and the black hole-stellar mass relation at $z=0$ \citep{Chaikin2026a}, rather than any luminosity-based constraints. The strong agreement with the observed LFs therefore suggests that the combination of the {\sc colibre} galaxy population and the present dust radiative transfer modelling captures the bulk of the relevant physics governing stellar emission, dust attenuation and dust re-emission in a broadly realistic manner. Although modest discrepancies remain in the MIR bands, likely reflecting the difficulty of modelling PAH emission, stochastic heating of small grains, the detailed heating of dust around star-forming regions and possibly missing components such as AGN and nebular continuum emission, the overall good agreement demonstrates that {\sc colibre} successfully reproduces the stellar population mix, the global dust reservoir and the associated energy balance of galaxies in the local Universe. The success is likely linked to several improvements in \colibre relative to earlier simulations:
\begin{enumerate}
\item {\it A more realistic treatment of the cold ISM and dust physics}. In contrast to previous large-volume simulations, such as EAGLE and IllustrisTNG, {\sc colibre} explicitly models the cold phase of the ISM, rather than relying on an effective equation of state \citep{Schaye2026,Ploeckinger2025}. In addition, dust is not treated as a purely passive component, but is coupled to H$_2$ formation, shielding, heating and cooling and metal depletion \citep{Ploeckinger2025,Trayford2026}. Rather than assuming a fixed dust mixture, {\sc colibre} follows the evolution of multiple grain species and grain sizes, including stellar dust production, grain growth by accretion, shattering, coagulation and sputtering \citep{Trayford2026}. This provides a more physical basis for radiative transfer modelling and likely contributes to the improved agreement across a broad wavelength range.

\item {\it A more realistic feedback model}. {\sc colibre} includes early stellar feedback, superbubble-like stellar feedback and AGN feedback within a more physically motivated and numerically controlled framework \citep{Schaye2026,Chaikin2023,Chaikin2026a,Benitez-Llambay2026}. These processes directly affect star formation histories, gas fractions and metal distributions and hence also the dust distribution.

\item {\it A self-consistent treatment of non-equilibrium chemistry and chemical enrichment}. {\sc colibre} models non-equilibrium chemistry and cooling, including the effects of shielding, cosmic rays and radiation fields \citep{Ploeckinger2025}, while also following detailed element-by-element chemical enrichment from multiple stellar channels \citep{Correa2026}. This improves the modelling of gas thermodynamics, metal enrichment and dust growth, all of which are important for recovering multi-band LFs.

\end{enumerate}

\section{Conclusions}
\label{sec:conclusion}
This is the first in a series of three papers, in which we aim comprehensively to study galaxy luminosity functions (LFs) from the FUV to the submillimetre and their redshift evolution as predicted by the new state-of-the-art hydrodynamical cosmological simulation suite, {\sc colibre} \citep{Schaye2026,Chaikin2026a}, combined with the calibration-free {\sc colibre-skirt} pipeline \citep{Gebek2026}, which is built upon the \skirt radiative transfer code \citep{Baes2011,Camps2015,Camps2020}. In this work, we studied the local ($z=0$) LFs from the FUV to the submillimetre and compared them with observations. Our main findings are:

\begin{enumerate}
\item \colibre simulations using different resolutions show good convergence in the correlations between stellar mass and galaxy properties such as total mass and SFR, likely owing to the synergy between stellar feedback by supernovae and early stellar feedback \citep{Benitez-Llambay2026}, which helps to regulate star formation as the resolution increases. However, the dust mass and cool gas mass exhibit resolution dependence, particularly at low masses. This is likely due to higher numerical resolution (lower particle masses) resolving better the formation of the cold and dense ISM phases, where dust growth via accretion is efficient. As a result, at fixed galaxy stellar mass, higher-resolution simulations can build up larger cold gas reservoirs and higher dust masses, leading to the observed resolution dependence (\autoref{fig:Mstar2Pros}). This effect is also seen in the correlation between galaxy stellar mass and broadband luminosities in the FUV (where dust attenuation is important) and IR (where dust emission is important) (\autoref{fig:mstar2lum}).

\item Using a representative subsample, we performed radiative transfer calculations with the calibration-free \cosk pipeline, including dust attenuation and emission and derived LFs from the FUV to the submillimetre for three \colibre simulations at different resolution and cube sizes (L025m5, L200m6 and L400m7). Across the full wavelength range, the LFs from the three simulations show excellent agreement from the FUV to the near-infrared bands and good agreement at the bright end at longer wavelengths, indicating convergence among \colibre simulations at different resolutions (Figs. \ref{fig:FUV2K} to \ref{fig:FIR}). This result is not in conflict with the somewhat poor convergence seen in the dust mass-stellar mass relation, because the bright end of the MIR to submillimetre LFs is dominated by galaxies with high dust masses, for which overall consistency is obtained.

\item The \cosk LFs match remarkably closely the observed LFs from the FUV to the $K$ band (\autoref{fig:FUV2K}) and the FIR to submillimetre bands ($70\,\micron$ to $850\,\micron$; \autoref{fig:FIR}). Considering that \colibre is only calibrated to reproduce the observed stellar mass function, the size-stellar mass relation and black hole mass-stellar mass relation at $z=0$, this agreement indicates that \colibre not only accurately recovers the stellar populations at $z=0$ but also predicts a reasonable representation of the cold dust reservoir and the total dust content in galaxies.

\item The \cosk LFs (with dust) show good agreement with observations down to the very faint end at $3.4\,\micron$, but predict galaxies that are not bright enough at the bright end from 8 to $24\,\micron$, leading to a truncated bright end compared to observations (\autoref{fig:MIR}). Our tests suggest that this discrepancy likely reflects a combination of modelling uncertainties, including the treatment of smoothing lengths, limited resolution, especially in star-forming regions, uncertainties in modelling the emission from PAHs and stochastically heated small grains, the absence of AGN and nebular continuum emission and the use of dust-free {\sc toddlers} templates for the star-forming regions. By contrast, our test indicates that dust self-heating has little effect on the MIR LF at $z=0$ and is therefore unlikely to explain the MIR discrepancy (see \autoref{sec:mismatch_mir} for a discussion).

\item By integrating the SEDs modelled with {\sc colibre-skirt} over $8-1000\,\micron$, we derived the total infrared (TIR) LF and compared it with observational estimates. The TIR LF agrees with observations at the faint end ($\log\,L_{\rm TIR}/\mathrm{L_{\odot}}\lesssim 11$) but is lower than observations at the bright end. The underprediction of the bright end of the TIR LF relative to some observational datasets may be due to a combination of underestimated MIR luminosities, redshift evolution in the observational samples, the omission of AGN-heated dust emission, the exclusion of dust self-heating in \cosk and possible cosmic/sample variance and limited wavelength coverage in some of the observations.
\end{enumerate}

\section*{Acknowledgements}
We thank the referee for their constructive comments, which improved the paper. We thank Jon Loveday for kindly providing his observed luminosity function data (\citealt{Loveday2012}) in a machine-readable format and Ana Vinterhalter for generously sharing the observational data she painstakingly compiled from the literature, which were used for comparisons in \citet{Trcka2020,Trcka2022}. We acknowledge support from STFC (ST/X001075/1). CSF acknowledges support from the European Research Council through Advanced Investigator grant DMIDAS (GA 786910). SB is supported by a UK Research and Innovation Future Leaders Fellowship (grant numbers MR/V023381/1 and UKRI2044). NA acknowledges financial support by the Flemish Fund for Scientific Research (FWO-Vlaanderen) through the research grant G0C4723N. ABL acknowledges support by the Italian Ministry for Universities (MUR) program “Dipartimenti di Eccellenza 2023-2027” within the Centro Bicocca di Cosmologia Quantitativa (BiCoQ) and support by UNIMIB’s Fondo Di Ateneo Quota Competitiva (project 2024-ATEQC-0050). EC acknowledges support from the Netherlands Organization for Scientific Research (NWO) through research programme Athena 184.034.002 and STFC consolidated grant ST/X001075/1. AD is supported by an STFC doctoral studentship. FH acknowledges funding from the Netherlands Organization for Scientific Research (NWO) through research programme Athena 184.034.002. SP acknowledges support by the Austrian Science Fund (FWF) through grant-DOI: 10.55776/V982. JT acknowledges support of a STFC Early Stage Research and Development grant (ST/X004651/1). This work used the DiRAC@Durham facility managed by the Institute for Computational Cosmology on behalf of the STFC DiRAC HPC Facility (\url{www.dirac.ac.uk}). The equipment was funded by BEIS capital funding via STFC capital grants ST/K00042X/1, ST/P002293/1, ST/R002371/1 and ST/S002502/1, Durham University and STFC operations grant ST/R000832/1. DiRAC is part of the UK National e-Infrastructure.

\section*{Data availability}
All the predicted \colibre luminosity functions at $z=0$ and $z=0.3$ in all bands, together with the observational datasets used for comparison, are provided as supplementary material and are available from the journal website and at \url{https://icc.dur.ac.uk/data/}. Readers are referred to \url{https://digastro.org} for additional datasets in machine-readable format.

\bibliographystyle{mnras}
\bibliography{ref} 

\appendix
\section{Effect of model set-up on luminosity functions}
\label{sec:many_effects}
In this section, we test the effects of various \colibre modelling choices and \skirt set-up options on the LFs across the entire wavelength range, from FUV to submillimetre.

\subsection{Effect of the AGN feedback model}
AGN can contribute additional heating, affecting both star formation and dust properties and consequently influencing LFs across the FUV to submillimetre wavelengths. In this work, we do not include any direct radiative contribution from AGN. Their impact on galaxy luminosities is accounted for only through their effects on the thermal and chemical state of the gas, and the resulting changes in, for example, star formation, metal enrichment and dust formation. Throughout the main body of the paper, we make use of the \colibre simulations with the purely thermal AGN feedback model. Thus, in this section, we test the impact of the AGN feedback model on the LFs from FUV to submillimetre. 

\begin{figure*}
\centering
\includegraphics[width=0.8\textwidth]{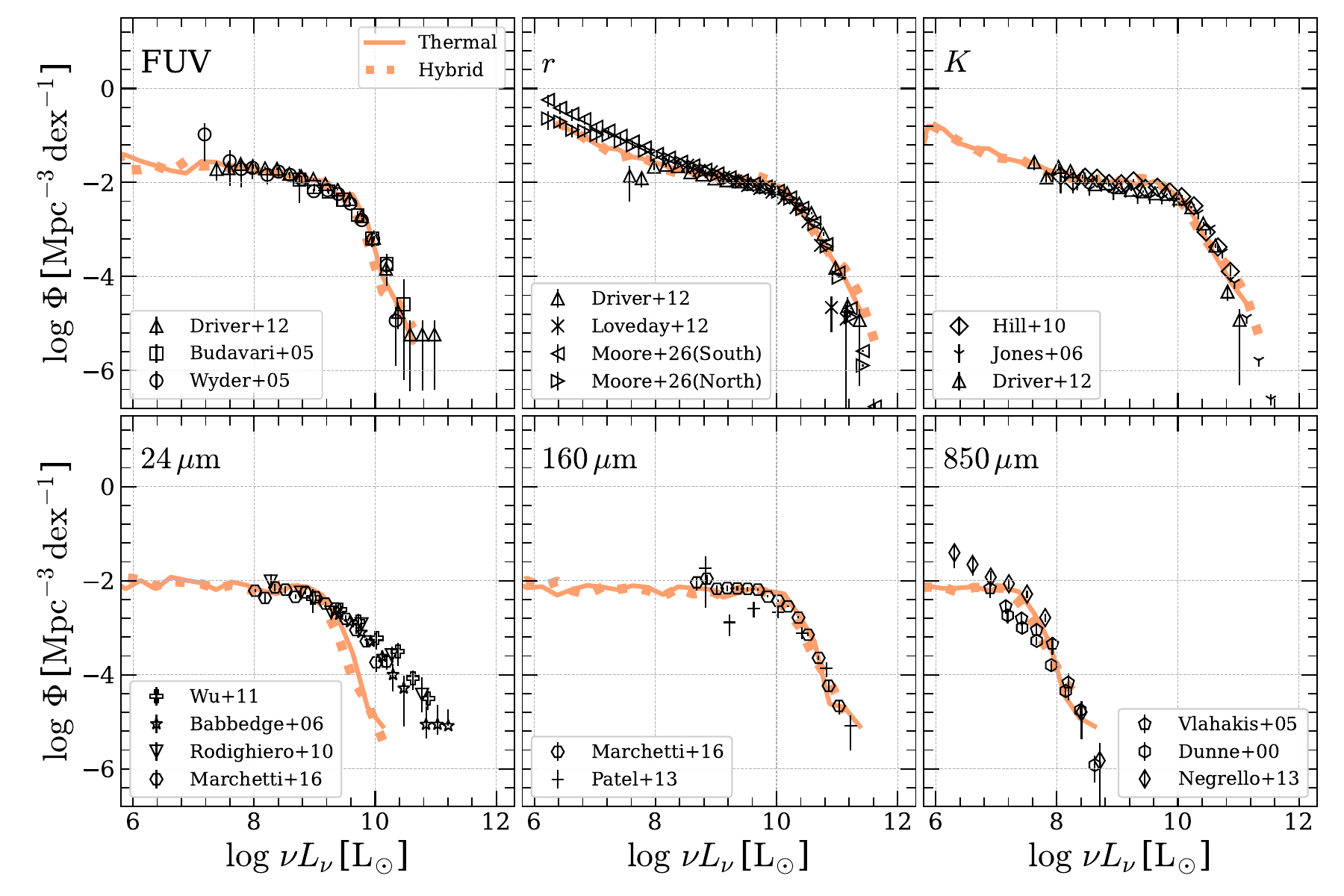}\vspace{-0.2cm}
\caption{Effect of the AGN feedback model on the dust-attenuated LFs in different bands. In each panel, the LF with the thermal AGN feedback model (the default in this paper) and that with the hybrid (thermal+jet) AGN feedback model (see \citealt{Husko2026} for details) are shown as solid and dotted curves, respectively. This test is carried out using simulations in a cube of size 100~cMpc at m6 resolution, L100m6 for the thermal AGN feedback model and L100m6h for the hybrid AGN feedback model (see \autoref{table:simulations} for details). No clear difference between the LFs obtained with the two AGN feedback models is seen.}
\label{fig:agn}
\end{figure*}

Specifically, we show in \autoref{fig:agn} the LFs derived from \colibre simulations with (1) the purely thermal AGN feedback model (the default setting in this work) and (2) the hybrid AGN feedback model, which includes jets \citep{Husko2026}, in six example bands covering the full wavelength range, including FUV (UV band), $r$ (optical), $K$ (near-infrared), 24 \micron\ (mid-infrared), 160 \micron\ (far-infrared) and 850 \micron\ (submillimetre). Here, we use the simulation at m6 resolution and in a cube of 100 cMpc to carry out this comparison (L100m6 and L100m6h; see \autoref{table:simulations} for details). Note that both AGN feedback models are calibrated to fit the same observational constraints at $z=0$, including the stellar mass function. As can be seen, across the entire wavelength range, the LFs based on the hybrid AGN feedback model are very similar to those based on the purely thermal AGN feedback model and also consistent with the observational data (except for the MIR bands as seen in \autoref{sec:mir}). This indicates that, for these calibrated AGN feedback models, the AGN feedback implementation does not significantly influence the LFs in any band at $z=0$.

\label{sec:agn}

\subsection{Effect of star-forming region resampling on LFs}
\label{sec:resampling}
\begin{figure*}
\centering
\includegraphics[width=0.8\textwidth]{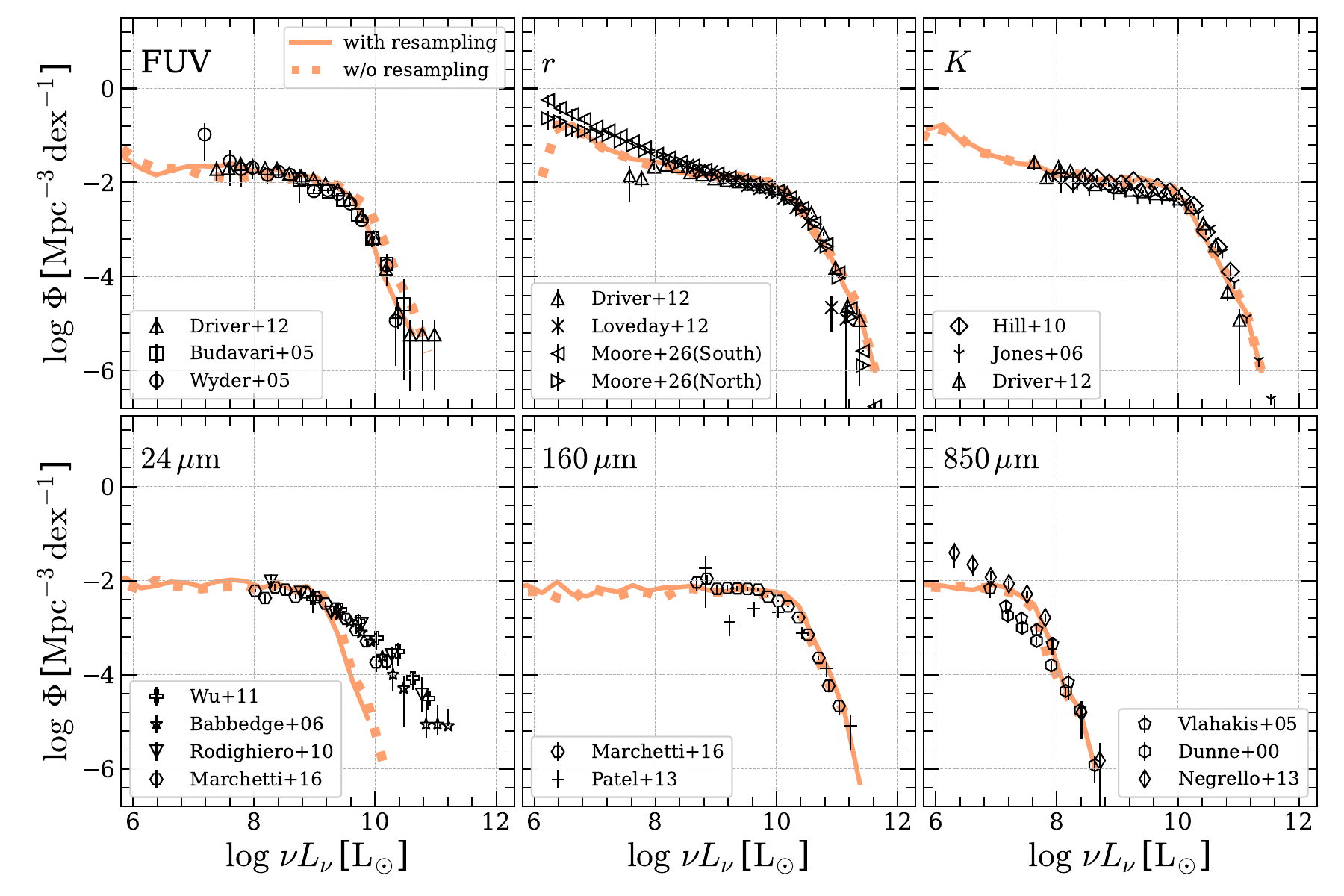}\vspace{-0.2cm}
\caption{Effect of star-forming region resampling (see \autoref{sec:skirt} for details) on the LFs (including dust) in different bands. In each panel, the LFs with and without resampling are shown as solid and dotted curves, respectively. This test is carried out with the \colibre simulation L200m6, with a cube size of $200\,\mathrm{cMpc}$ and a mass resolution at the m6 level (see \autoref{table:simulations}).}
\label{fig:sf_resampling}
\end{figure*}

As mentioned in \autoref{sec:skirt}, due to the limited resolution of the simulations, we resample the young stellar particles using the star-forming gas, assuming a constant SFR over the past 10~Myr. In \autoref{fig:sf_resampling}, we show the comparison between LFs with and without star-forming region resampling in six example bands covering the full wavelength range. As shown, the resampling of star-forming regions affects only the FUV LF, shifting it towards brighter luminosities by up to 0.4 dex. In the other bands, the resampling has no noticeable effect.

\subsection{Effect of aperture size}
\label{sec:aperture}
Throughout the paper, we adopt a projected circular aperture with a proper radius of 50~kpc and include only gravitationally bound particles, thereby excluding satellites. In observations, source-extraction apertures, rather than fixed physical apertures, are typically used to recover the total luminosities of galaxies as accurately as possible while minimizing contamination from neighbouring sources, the sky background and other observational effects. These apertures are defined on the basis of the observed surface-brightness distributions of galaxies; for example, \citet{Driver2012} adopted Kron apertures, whereas \citet{Loveday2012} used Petrosian apertures. In our analysis, because only particles gravitationally bound to each galaxy are included, adopting a large aperture does not introduce contamination from nearby galaxies. \citet{Trcka2022} showed that a 30~kpc aperture provides a good approximation to commonly used Petrosian apertures for massive galaxies, although even such an aperture may still miss some light from the most extended systems.

\begin{figure*}
\centering
\includegraphics[width=0.8\textwidth]{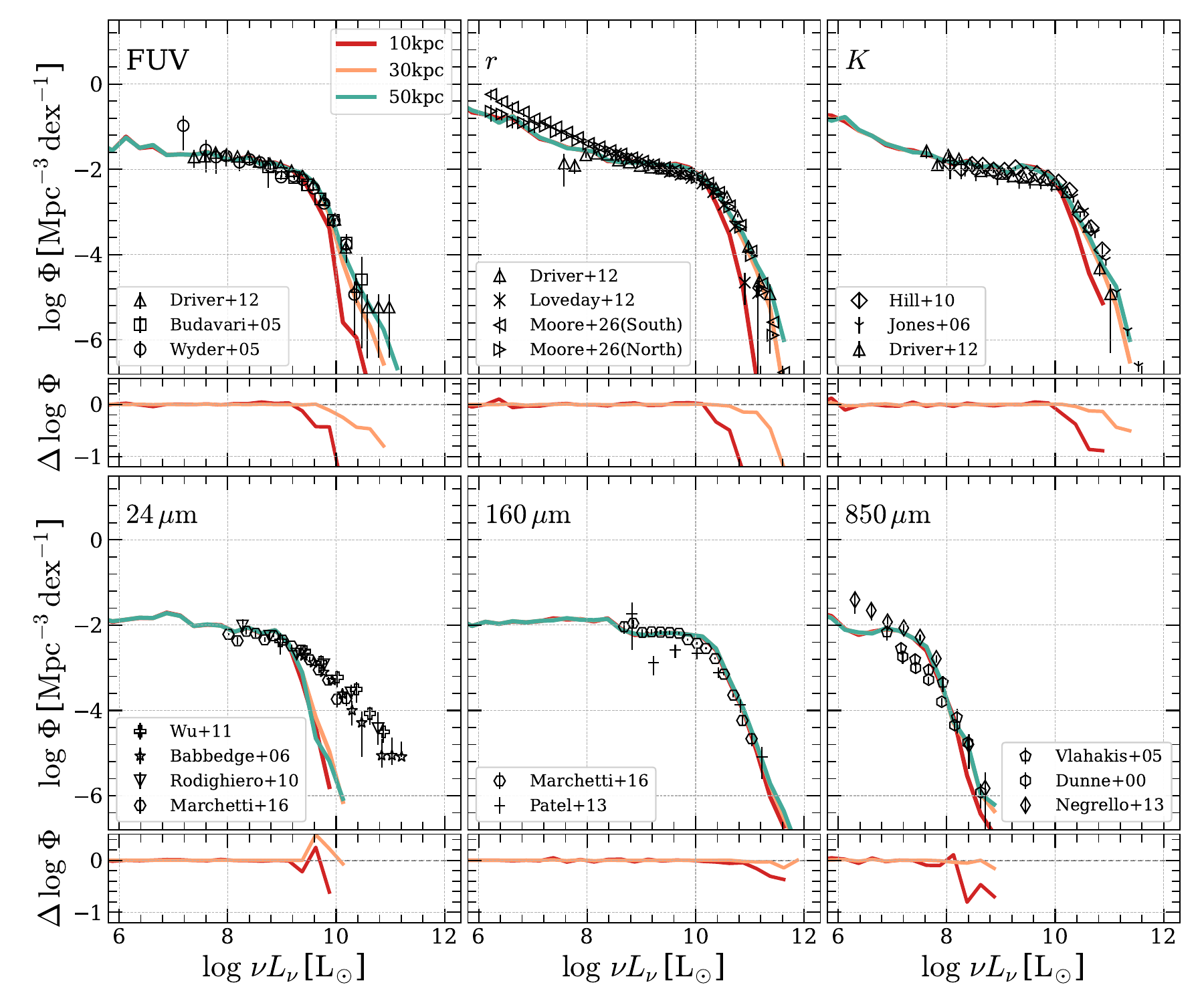}\vspace{-0.2cm}
\caption{Effect of aperture sizes on the LFs (including dust) in different bands. In each panel, the LFs with different aperture sizes are shown with different colours. At the bottom of each panel, we show the differences in galaxy abundance between the 10 and 30 kpc apertures and the fiducial 50 kpc aperture, defined as $\Delta \log \Phi = \log\,\Phi_{\rm apt} - \log\,\Phi_{\rm 50\,kpc}$, where $\log\,\Phi_{\rm apt}$ denotes the abundance measured using the 10 or 30 kpc aperture, as a function of $\log\,\nu L_{\nu}$. All \cosk LFs presented here are the combined results of three \colibre simulations with different resolutions (see \autoref{sec:result} for details of the procedure used to combine them).}
\label{fig:aperture}
\end{figure*}

In \autoref{fig:aperture}, we compare the LFs in six representative bands obtained using aperture radii of 10, 30 and the fiducial 50 kpc. In the FUV, $r$ and $K$ bands, the 10 kpc aperture produces substantially lower bright-end abundances than the larger apertures. Relative to the fiducial 50 kpc aperture, the 30 kpc aperture reduces the abundance in the brightest bins by up to approximately $0.8$, $1.0$ and $0.5$ dex in the FUV, $r$ and $K$ bands, respectively. These relatively large vertical differences in number density arise primarily from the steep slopes of the LFs at the bright end. At a fixed number density of $\log\,\Phi/\mathrm{Mpc^{-3}\,dex^{-1}}=-4$, the corresponding horizontal shifts in galaxy luminosity are negligible, while at $\log\,\Phi/\mathrm{Mpc^{-3}\,dex^{-1}}=-6$, they are approximately $0.3$, $0.2$ and $0.1$ dex in the FUV, $r$ and $K$ bands, respectively. The LFs obtained using the 30~kpc aperture therefore remain broadly consistent with those obtained using the fiducial 50~kpc aperture. By contrast, the mid- and far-infrared LFs are nearly insensitive to the adopted aperture.

\section{Effect of random sampling on luminosity functions}
\label{sec:random_sampling}
As mentioned in \autoref{sec:sample}, we do not perform radiative transfer on all galaxies in the \colibre simulation cubes of interest due to computational cost, but instead model a representative subsample. Here, we assess the impact of this random sampling strategy on the LFs. We first construct a subsample from the standard sample (the originally selected sample; see \autoref{sec:sample}) using the following procedure: for the originally selected galaxies in a given stellar mass-SFR bin (with bin sizes of 0.5 dex in both stellar mass and SFR), if the number of galaxies exceeds half of the original target number, 25 for m5 and 15 for m6 and m7, we randomly select half of the original target number of galaxies; if it is fewer than half of the original target number, we include all originally selected galaxies. The sampling ratios are reduced accordingly and the new LFs are re-calculated with \autoref{eq:lf}. This new, smaller subsample is referred to as the \textit{reduced subsample}, relative to the \textit{standard subsample} listed in \autoref{table:simulations}. 

\begin{figure*}
\centering
\includegraphics[width=0.9\textwidth]{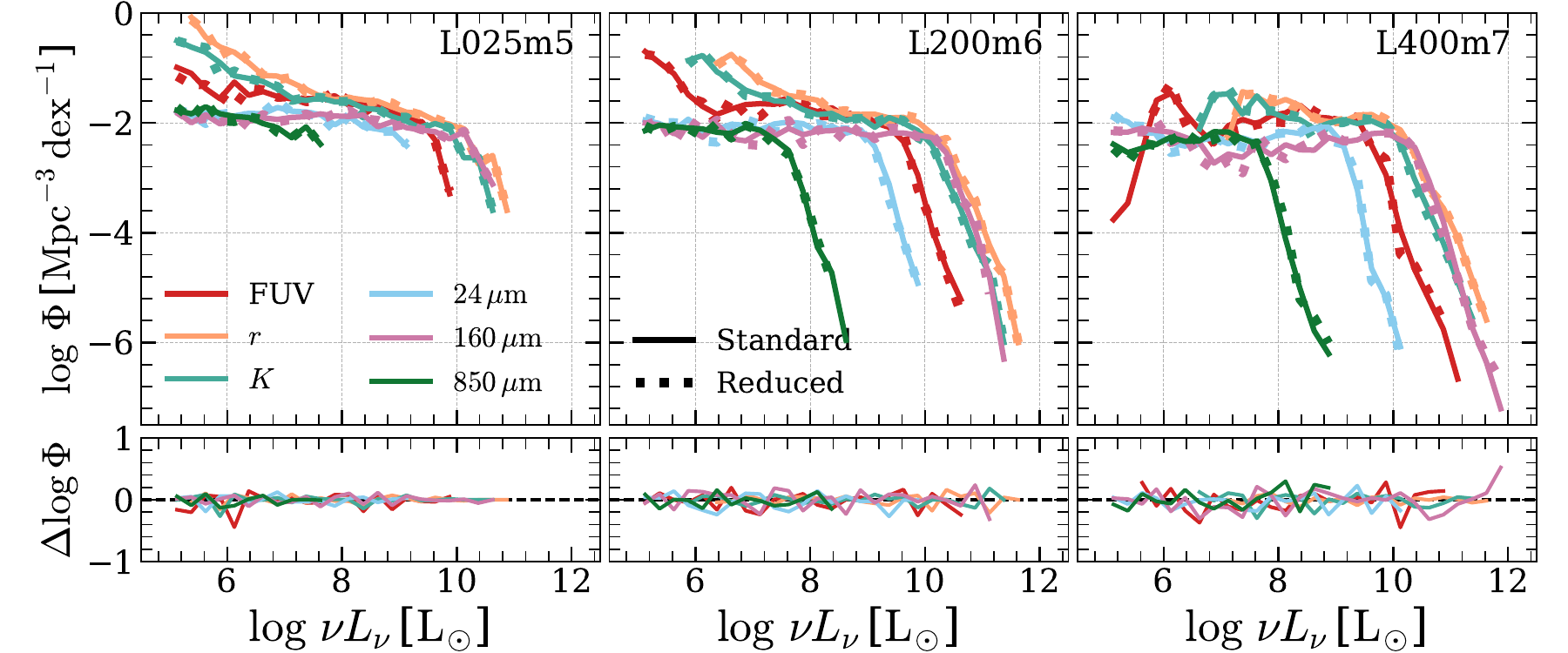}\vspace{-0.2cm}
\caption{Effect of random sampling on example dust-attenuated LFs. Results for different \colibre simulations are shown in separate panels (from left to right: L025m5, L200m6 and L400m7). In each panel, the LFs in different bands are indicated by different colours. The solid curves represent the LFs of the {\it standard} subsample (the originally selected subsample; see \autoref{sec:sample} and \autoref{table:simulations}), while the dotted curves correspond to the LFs of the {\it reduced} subsample (with smaller sample sizes; see Appendix~\ref{sec:random_sampling}). The lower row shows the number density difference, $\Delta \log\Phi \equiv \log\Phi_{\rm standard}-\log\Phi_{\rm reduced}$, as a function of $\log \nu L_{\nu}$. The differences are generally small, indicating that the current sampling method yields accurate LF estimates for the whole simulation volume.}
\label{fig:subsampling}
\end{figure*}

In \autoref{fig:subsampling}, we compare the LFs derived from the standard and reduced subsamples across the three \colibre simulations with different resolutions in several example bands spanning short to long wavelengths. As can be seen, across all bands, the LFs based on the reduced sample closely match those from the standard sample, indicating that even a smaller sample is sufficient to accurately recover the LFs of the full galaxy population, thereby demonstrating the robustness and representativeness of our sampling method. In this way, we save over 95\% of the computational time while still deriving accurate LFs.

\section{Monochromatic luminosity calculation}
\label{sec:mono_lum_appendix}
\begin{figure*}
\centering
\includegraphics[width=0.9\textwidth]{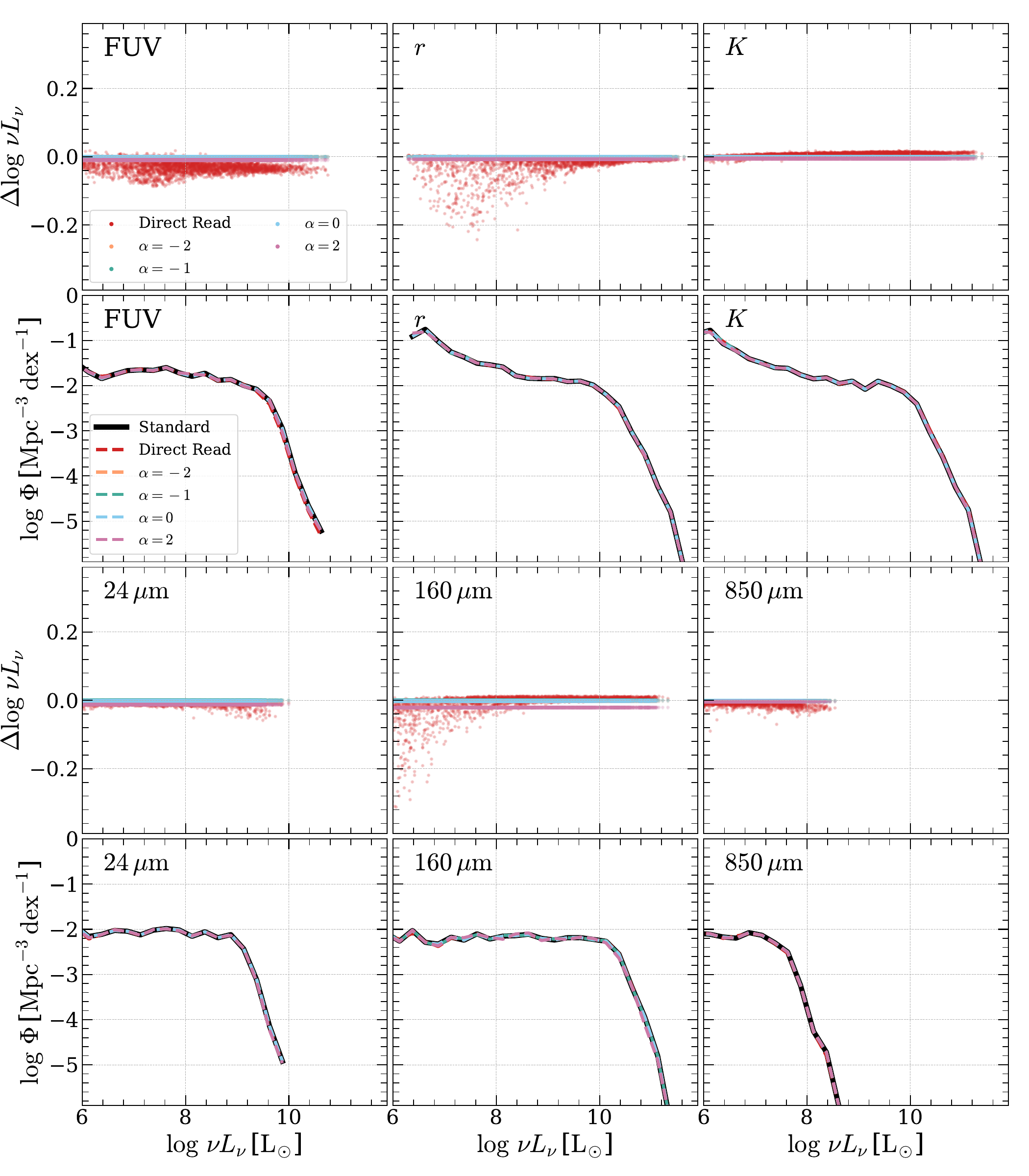}\vspace{-0.2cm}
\caption{Comparison of different methods for obtaining monochromatic luminosities. We show the results across bands from FUV to submillimetre. For each band, the upper panel shows the relation between the {\it standard} $\log\,\nu L_{\nu}$ (the filter-integrated flux) and the difference between the standard $\log\,\nu L_{\nu}$ and monochromatic luminosities obtained using other methods, while the lower panel shows the comparison of LFs derived from these methods. Here, we compare the filter-integrated monochromatic luminosities to those obtained with five other approaches: (1) directly reading from SEDs and (2-5) assuming a constant spectral index $\alpha$ ($F_{\nu}\propto \nu^{\alpha}$) with $\alpha = -2, -1, 0, 2$.}
\label{fig:monolum}
\end{figure*}

Monochromatic luminosities can be calculated using different methods (see \autoref{sec:mono_lum}). In this work, we follow the practice of earlier studies investigating the LFs of simulations using \skirt (e.g. \citealt{Trcka2020,Gebek2024}) and adopt the filter-averaged flux as the effective monochromatic flux. In this section, we also investigate the performance of other methods for calculating monochromatic luminosities, including (1) directly reading the flux from the SEDs and (2-5) assuming a constant spectral index ($F_{\nu}\propto \nu^{\alpha}$) with $\alpha = -2, -1, 0, 2$. In \autoref{fig:monolum}, we present (1) the differences in luminosities (in the format of $\nu L_{\nu}$) for individual galaxies and (2) the corresponding differences in LFs. We show results for six example bands (FUV, $r$, $K$, 24, 160 and 850 \micron).

As shown in the figure, for most of the selected bands (i.e., FUV, $r$, $K$, 160 \micron and 850 \micron), although the monochromatic luminosities from different methods for individual galaxies show slight differences (especially those directly read from the SEDs, i.e., red dots), the LFs show remarkably good agreement across different methods. Given that the filter-averaged luminosity is well defined, does not rely on any assumptions about the SED shape and is directly comparable to observational measurements, we adopt the filter-averaged luminosity to calculate LFs in this work.

\section{Effect of redshift evolution on luminosity functions}
\label{sec:LF_z-dependence}
As mentioned in \autoref{sec:obs_data} (see also \autoref{table:filter}), observational studies typically use objects within a small redshift range (e.g. $z=0-0.3$) to calculate the so-called ``local'' LFs, instead of using galaxies exactly at $z= 0$, while in our work, we only use galaxies at $z=0$ from {\sc colibre}. This introduces a potential risk if the LF evolves significantly even at very low redshifts. Thus, in this section, we carry out an investigation of the redshift dependence of \cosk LFs from $z=0$ to $z=0.3$ (the upper redshift limit of most observational local LF studies). To do so, we select a subsample of galaxies at $z=0.3$ from L025m5, L200m6 and L400m7, following the same selection process described in \autoref{sec:sample}, and calculate the LFs for six example bands (FUV, $r$, $K$, 24, 160 and 850 \micron) accordingly. 

\begin{figure*}
\centering
\includegraphics[width=0.8\textwidth]{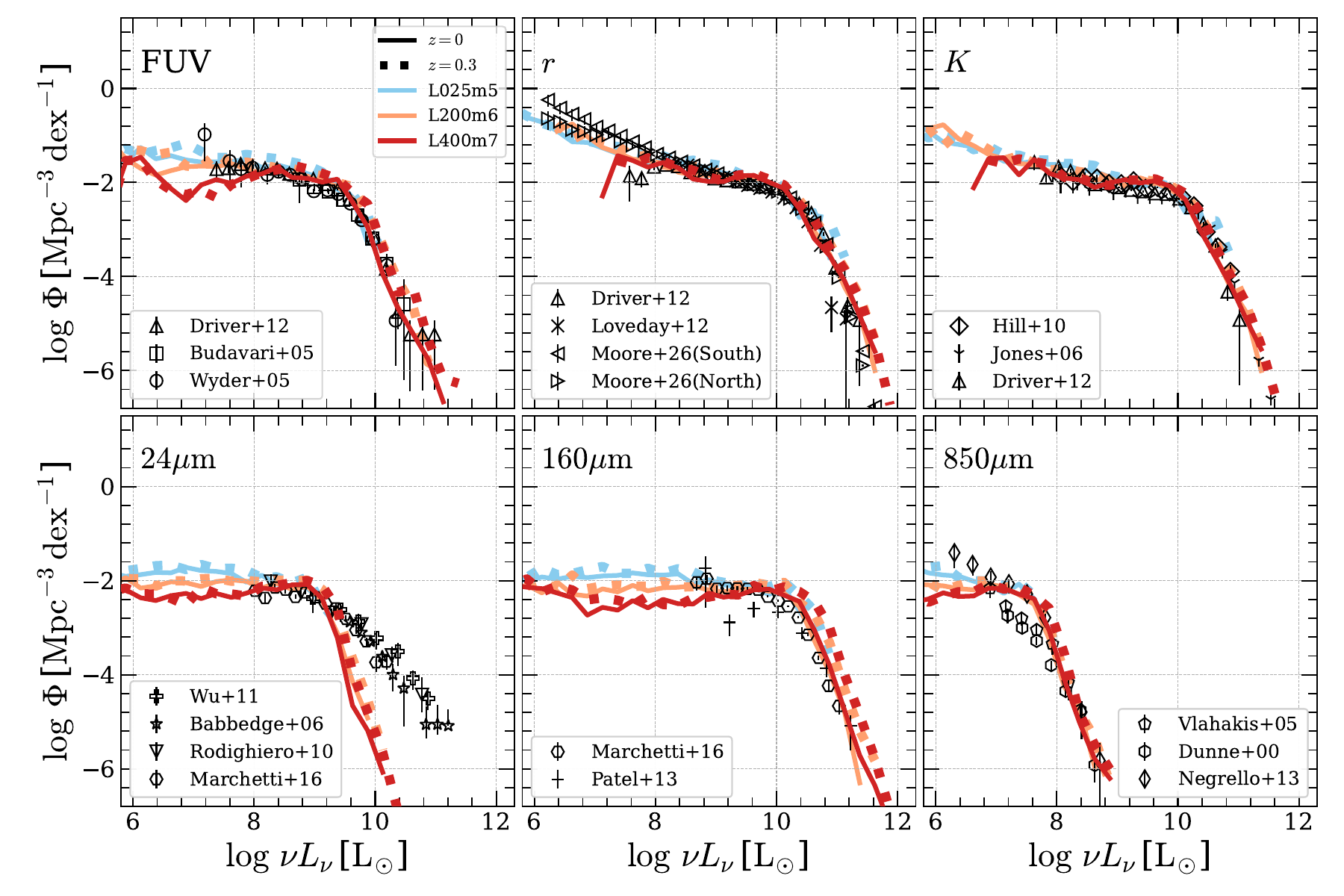}\vspace{0cm}
\caption{Comparison of LFs (including dust) at $z=0$ (solid curves) and $z=0.3$ (dotted curves) across bands from FUV to submillimetre. Results from different simulations are shown in different colours. The $z=0.3$ LFs are shifted towards the bright end relative to the $z=0$ LFs, with the offset being larger in the FUV, MIR (24 \micron) and FIR (160 \micron) bands.}
\label{fig:zdependence}
\end{figure*}

In \autoref{fig:zdependence}, we present a comparison of \cosk LFs at $z=0$ and $z=0.3$. As can be seen, the $z=0.3$ \cosk LFs are shifted towards higher luminosities relative to the $z=0$ \cosk LFs in all bands for all simulations, with the difference being most pronounced in the FUV, MIR (24 \micron) and FIR (160 \micron) bands. Specifically, $z=0.3$ galaxies are $\approx 0.2\,\rm dex$ more luminous than $z=0$ galaxies in the FUV band at a number density of $10^{-4}\,\mathrm{Mpc^{-3}\,dex^{-1}}$ and are $\approx 0.4\,\rm dex$ more luminous in the FIR (160 \micron) band at the same number density. This is consistent with observational studies showing relatively mild evolution of the FUV LF at low redshift (e.g. \citealt{Arnouts2005}), but stronger luminosity evolution in the MIR/FIR LFs \citep[e.g.][]{LeFloch2005,Gruppioni2013,Patel2013}. The three simulations predict very similar changes in the LFs from $z=0$ to $z=0.3$, indicating that the inferred redshift evolution is insensitive to numerical resolution over the luminosity ranges where the simulations overlap. Given that the observational data used for comparison in this work include both very local samples ($z\approx 0$) and samples that extend to $z\approx 0.3$, we use only the galaxies at $z=0$ to calculate the ``local'' LFs in the main body of this work. 

\label{lastpage}
\end{document}